\documentclass[reprint,twocolumn]{revtex4}
\usepackage{amsmath} 
\usepackage{amsfonts} 
\usepackage{graphicx} 
\usepackage{hyperref}
\usepackage{stix}

\begin{document}
\title{Persistence of power-law correlations in nonequilibrium steady states of gapped quantum spin chains}
\author{Jarrett L. Lancaster}
\email{jlancas2@highpoint.edu}
\author{Joseph P. Godoy}
\altaffiliation{Currently at Appalachian State University, Department of Physics and Astronomy, 525 Rivers Street, Boone, NC 28608} 

\affiliation{Department of Physics, High Point University, One University Parkway, High Point, NC 27262}

\date{\today}

\begin{abstract}
The existence of quasi-long range order is demonstrated in nonequilibrium steady states in isotropic $XY$ spin chains including of two types of additional terms that each generate a gap in the energy spectrum. The system is driven out of equilibrium by initializing a domain-wall magnetization profile through application of an external magnetic field and switching off the magnetic field at the same time the energy gap is activated. An energy gap is produced by either applying a staggered magnetic field in the $z$ direction or introducing a modulation to the $XY$ coupling. The magnetization, spin current, and spin-spin correlation functions are computed analytically in the thermodynamic limit at long times after the quench. For both types of systems, we find the persistence of power-law correlations despite the ground-state correlation functions exhibiting exponential decay. 

\end{abstract}

\maketitle
\section{Introduction}
Many-body dynamics in closed-quantum systems has exploded in popularity as an area of intense research over the last decade~\cite{Silva2011,Cazalilla2011,Eisert2015,Borgonovi2016,Mitra2018}. The surge in popularity of low-dimensional quantum dynamics as a theoretical area of study has largely been spurred by remarkable advances in the experimental simulation of low-dimensional systems with tightly controlled and highly tunable interactions. With simulations of toy-model-like systems readily available in the form of carefully designed setups which manipulate ultracold atoms in optical traps~\cite{Weiss2006,Greiner2011,Senko2015,Bloch2015,Labuhn2016,Rajagopal2017,Naini2018}, it is possible to explore questions of thermalization and relaxation experimentally using models which are simple enough to allow for a thorough analytic treatment. 

Perhaps the most well-known class of such theoretically simple systems is that of quantum spin chains, which consist of a one-dimensional lattice of spin degrees of freedom. Bethe first solved~\cite{Bethe1931} the Heisenberg spin chain employing a method which would become known as the ``Bethe ansatz,'' which has since been developed into a framework suitable for investigating the exact dynamics of integrable systems away from equilibrium~\cite{Mossel2010}. The simpler $XY$ model was introduced and shown to map to a system of free fermions by Lieb, Shultz, and Mattis~\cite{LSM1961}. Due to the simplicity of the basic $XY$ model, it has become a popular system for investigating nonequilibrium physics analytically. The $XY$ spin chain with an external magnetic field leads to a surprisingly rich phase diagram, and its correlation functions were explored quite thoroughly in the work of Barouch, McCoy and collaborators~\cite{Barouch1,Barouch2,Barouch3,Barouch4}.

It is noteworthy that Ref.~\onlinecite{Barouch1} devotes a section to computing the dynamics of observables following an abrupt change in the external magnetic field---an early example of a ``quantum quench'' protocol~\cite{Mitra2018}. In recent years, the quantum quench has become an extremely popular device for generating nonequilibrium dynamics which can be simulated experimentally through experiments with ultracold atoms with rapidly tunable external fields~\cite{Bloch2012,Wen2013}. Experimentally, tuning the parameters of the initial system allows one to probe the dynamics after the quench for a wide variety of carefully chosen initial states. 

From the theorist's point of view, a quantum quench is essentially a formal procedure for investigating time evolution with an arbitrary initial state. After a sudden change in system parameters, the ground state of the initial eigenstate is no longer the ground state of the Hamiltonian which generates time evolution. Very often, the initial state is no longer any eigenstate of the final Hamiltonian. While generic, nonintegrable systems coupled to an external environment would be expected to thermalize, the situation is more subtle for isolated many-body systems. The eigenstate thermalization hypothesis provides a general mechanism by which expectation values of observables in isolated, nonintegrable systems can approach thermal values despite the absence of coupling to an external environment~\cite{Deutsch1991,Srednicki1994,Deutsch2018}. 

For integrable systems, which contain a conserved quantity for every degree of freedom, the dynamics is tightly constrained. Consequently, an integrable system generally cannot relax fully to thermal equilibrium in the traditional sense and instead relaxes to a generalized Gibbs ensemble (GGE)~\cite{Rigol2006,Rigol2007,Vidmar2016}. The crossover between integrable and non-integrable systems results in a smooth transition~\cite{Rigol2009}. For systems in which integrability is only weakly broken, the system often relaxes to a long-lived GGE before eventually thermalizing at very long times~\cite{Kollar2011,Howell2019}. Even for systems which can be mapped to noninteracting quasiparticles, observables can exhibit nontrivial dynamics as a consequence of dephasing of the system's eigenmodes~\cite{Barthel2008}. Moreover, operators which are nonlocal with respect to the quasiparticles can lead to effectively thermal behavior~\cite{Rossini2011} despite the trivially integrable nature of the noninteracting system. 

The majority of studies of quench dynamics focus on spatially homogeneous systems. However, spatial inhomogeneities in the initial state provide particularly simple means of generating nonequilibrium dynamics. Experimentally, domain-wall magnetization configurations can be created by application of a spatially varying magnetic field~\cite{Weld2009,Weld2011}. In the isotropic $XY$ spin chain, a linearly varying magnetic field can be used to create a domain-wall magnetization profile. By suddenly switching off the magnetic field, the domain wall is observed to spread ballistically~\cite{Antal1999}. However, evidence of the system being far from equilibrium is found in the athermal relaxation of the transverse spin-spin correlation function, which acquires oscillations at the scale of the lattice~\cite{Lancaster2010a}. The isotropic $XY$ model may be recast in terms of hardcore bosons, in which these oscillations correspond to quasicondensates of the bosonic modes~\cite{Rigol2004,Rigol2006}. In the spin language, the oscillations are directly related to a spin current~\cite{Antal1997,Antal1998} which emerges as magnetization is being transported across the domain wall as the profile spreads. It has been shown recently that the time evolution of the domain-wall initial state is well described as an eigenstate of an emergent, effective Hamiltonian~\cite{Vidmar2017,Vidmar2017a}.

Domain-wall dynamics have also been investigated in the anisotropic $XY$ model~\cite{Aschbacher2003,Aschbacher2006,KormosSciPost,EislerSciPost,Lancaster2016,EislerPRB}. Like the isotropic $XY$ model, the anisotropic model maps to free fermions so that the dynamics may be computed exactly. Aside from exact diagonalization methods~\cite{Santos2011}, domain walls in the interacting $XXZ$ model have been treated numerically using time-dependent density matrix renormalization group ($t$DMRG) methods~\cite{Gobert2004,Sabetta2013} and theoretically using a bosonization approach~\cite{Lancaster2010a}. More recently, generalized hydrodynamics~\cite{ColluraPRB,FagottiPRB} has been employed to obtain analytic results which agree well with numerical calculations. 

With some notable exceptions~\cite{Foster2011,CalabreseSciPost}, detailed investigations of quench dynamics in which translation-invariance is broken by the Hamiltonian generating time evolution are scarce in the literature. Breaking translation invariance in spin chain models is most easily accomplished by adding a position dependence to external magnetic fields or nearest-neighbor couplings. In the present work, we address both types of terms. 

All models considered are extensions of the isotropic $XY$ model as described by the Hamiltonian~\cite{LSM1961}
\begin{eqnarray}
\hat{H}_{xy} & = & -J\sum_{j}\left[\hat{S}_{j}^{x}\hat{S}_{j+1}^{x} + \hat{S}_{j}^{y}\hat{S}_{j+1}^{y}\right].\label{eq:hxy}
\end{eqnarray}
In this paper we break translation invariance by performing each of the following two modifications to Eq.~(\ref{eq:hxy}). First, we add a term corresponding to a ``staggered'' magnetic field which oscillates at the scale of the lattice
\begin{eqnarray}
\hat{H}_{s} & = & \hat{H}_{xy} + m\sum_{j}(-1)^{j}\hat{S}_{j}^{z},
\end{eqnarray}
where $m>0$ is a constant. The second type of modification is to modulate the nearest-neighbor, $XY$ coupling $J\rightarrow J_{j} = J\left(1-(-1)^{j}\delta\right)$ for $0<\delta <1$. Such modulation leads to dimerization in which an energy gap appears in the spectrum~\cite{Pincus1971}. The addition of a staggered magnetic field also leads to the appearance of an energy gap, and this type of perturbation has been used previously~\cite{Foster2011} as a particularly simple mechanism to generate a gap in the energy spectrum of free fermions confined to a lattice. Additionally, a self-consistent version of the staggered field perturbation has recently been used to investigate the dynamics of a Bose-Hubbard model with a particular form of global-range interactions in the thermodynamic limit, where the system maps to a free-fermion model with a self-consistency condition~\cite{Rieger2018,Rieger2018b}. 

The work presented in this article focuses on the long-time behavior of one-body observables and correlation functions in the nonequilibrium steady state that forms in the center of the system as the domain wall broadens, or ``melts.'' Previous efforts have demonstrated that the central subsystem relaxes to a GGE-like steady state when time evolution is generated by the isotropic~\cite{Sabetta2013} and anisotropic~\cite{Lancaster2016} $XY$ models. In these cases, an effective momentum distribution describing the long-time limit of the central subsystem is obtained by expanding the initial momentum correlation matrix $\left\langle c_{p+\frac{q}{2}}^{\dagger}c_{p-\frac{q}{2}}\right\rangle$ for small $q$~\cite{Sabetta2013,Racz2010}. As these models map to free fermions, all observables can be computed from this effective momentum distribution. In this paper we extend the analysis to dynamics generated by the two gapped models discussed above. Due to the broken translation invariance, the effective momentum distribution is replaced by an effective Wigner function which inherits an explicit position dependence. 

Despite the presence of an energy gap in the spectrum of the Hamiltonian which generates time evolution, we ultimately find the persistence of power-law correlations in the spin-spin correlation functions. The existence of power-law correlation functions in a gapped model far from equilibrium has been demonstrated previously in continuum systems~\cite{Iucci2010,Lancaster2010b} and more recently in lattice systems~\cite{Rieger2018b}. One noteworthy feature of our results is that the power-law correlations depend crucially on the domain-wall initial state. As we will discuss, a homogeneous quench from the ground state of the $XY$ model results in at least one correlation function decaying exponentially with distance, similarly to its behavior in the ground state of the gapped model.

The paper is organized as follows. Section~\ref{sec:gs} contains a summary of the ground state properties of the models considered to establish a baseline to which the nonequilibrium results may be compared. The main results for the long-time limit of observables in both types of gapped systems are derived in Section~\ref{sec:dynamics}. Also included in this section is a brief investigation of quenches in which the initial state does not possess a domain wall magnetization profile and a comparison of our results to domain-wall dynamics in the gapped anisotropic $XY$ model. Lastly, Section~\ref{sec:conc} contains a brief discussion and outlook on future work.

\section{Ground state observables}
\label{sec:gs}
This section is devoted to a brief exploration of the ground-state properties of the models considered in the remainder of the paper. In the process of obtaining the low-energy physics, the basic nomenclature used throughout the paper is introduced. This section also contains the definitions of the particular observables which will be computed away from equilibrium in Sec.~\ref{sec:dynamics}.
\subsection{Staggered field}
The isotropic $XY$ model with a staggered (alternating) magnetic field is described by the following Hamiltonian,
\begin{eqnarray}
\hat{H}_{s} & = & -J\sum_{j}\left[\hat{S}_{j}^{x}\hat{S}_{j+1}^{x} + \hat{S}_{j}^{y}\hat{S}_{j+1}^{y}\right] + m\sum_{j}(-1)^{j}\hat{S}_{j}^{z}.\label{eq:hs}
\end{eqnarray}
We employ units in which $\hbar = 1$ and the lattice spacing $a\rightarrow 1$. Previous work has investigated both static~\cite{Igloi1988,Fel'dman2005} and dynamical~\cite{Crooks2007,Deng2008,Jafari2019} properties of the isotropic $XY$ model in the presence of a staggered magnetic field. The Jordan-Wigner transformation~\cite{JordanWigner,LSM1961} may be employed to write the spin operators in terms of spinless fermions, 
\begin{eqnarray}
\hat{S}_{j}^{+} & = & c_{j}^{\dagger}\exp\left[ i\pi\sum_{n=1}^{j-1}c_{n}^{\dagger}c_{n}\right],\label{eq:jw1}\\
\hat{S}_{j}^{-} & = & \exp\left[ -i\pi\sum_{n=1}^{j-1}c_{n}^{\dagger}c_{n}\right]c_{j},\label{eq:jw2}\\
\hat{S}_{j}^{z} & = & c_{j}^{\dagger}c_{j} - \frac{1}{2},\label{eq:jw3}
\end{eqnarray}
where $\hat{S}^{\pm}_{j} \equiv \hat{S}^{x}_{j} \pm i\hat{S}_{j}^{y}$. Up to an irrelevant constant, the transformation in Eqs.~(\ref{eq:jw1})--(\ref{eq:jw3}) can be used to recast Eq.~(\ref{eq:hs}) as
\begin{eqnarray}
\hat{H}_{s} & = & -\frac{J}{2}\sum_{j}\left[c_{j}^{\dagger}c_{j+1} + c_{j+1}^{\dagger}c_{j}\right] + m\sum_{j}(-1)^{j}c_{j}^{\dagger}c_{j}.\label{eq:hstag}
\end{eqnarray}
One may diagonalize $\hat{H}_{s}$ in terms of quasiparticles $\gamma_{k}$ using the following Bogoliubov transformation~\cite{Iucci2010,Foster2011}
\begin{eqnarray}
\left(\begin{array}{c} c_{k}\\ c_{k+\pi}\end{array}\right) & = & \left(\begin{array}{cc} \cos\frac{\theta_{k}}{2} & \sin\frac{\theta_{k}}{2}\\ -\sin\frac{\theta_{k}}{2} & \cos\frac{\theta_{k}}{2}\end{array}\right)\left(\begin{array}{c} \gamma_{k} \\ \gamma_{k+\pi}\end{array}\right),\label{eq:bog1}
\end{eqnarray}
valid for $|k|<\frac{\pi}{2}$. Here the Fourier transform of the $c_{j}$ is given explicitly by
\begin{eqnarray}
c_{k} & = & \frac{1}{\sqrt{L}}\sum_{j}e^{-ikj}c_{j},
\end{eqnarray}
where $L =Na$ is the size of the system and $N$ is the number of lattice sites. We work in the thermodynamic limit where we take $N\rightarrow \infty$ in all results. Using Eq.~(\ref{eq:bog1}) in Eq.~(\ref{eq:hstag}) results in a diagonal Hamiltonian of the form
\begin{eqnarray}
\hat{H}_{s} & = & -\sum_{|k|<\frac{\pi}{2}}\epsilon_{k}\left[\gamma_{k}^{\dagger}\gamma_{k} - \gamma_{k+\pi}^{\dagger}\gamma_{k+\pi}\right],\label{eq:hstag1}
\end{eqnarray}
for the specific choice
\begin{eqnarray}
\tan\theta_{k} & = & \frac{m}{J\cos k},\label{eq:tan1}
\end{eqnarray}
so that $\epsilon_{k} = \sqrt{\left(J\cos k\right)^{2} + m^{2}}$. The staggered magnetic field, which acts as a spatially-varying chemical potential in the fermion language, thus doubles the size of the unit cell, resulting in the first Brillouin zone being cut in half. Accordingly, the free fermion dispersion splits into upper and lower bands with a gap opening at the Fermi points, $k= \pm\frac{\pi}{2}$ of magnitude $\Delta \epsilon = 2m$. The ground state $\left|\Phi_{0}\right\rangle$ consists of a fully-filled band containing all quasiparticles in the lower branch of the dispersion, for which $\epsilon_{k}<0$, 
\begin{eqnarray}
\left|\Phi_{0}\right\rangle & = & \prod_{|k|<\frac{\pi}{2}}\gamma_{k}^{\dagger}\left|0\right\rangle.
\end{eqnarray}
Employing Wick's theorem, observables such as magnetization and correlation functions can be constructed from the basic correlations $\left\langle c_{j}^{\dagger}c_{j+n}\right\rangle = \left\langle \Phi_{0}\right| c_{j}^{\dagger}c_{j+n}\left|\Phi_{0}\right\rangle$. Transforming to momentum space,
\begin{eqnarray}
c_{j} & = & \frac{1}{\sqrt{N}}\sum_{|k|<\frac{\pi}{2}}e^{ikj}\left[c_{k} + (-1)^{j}c_{k+\pi}\right],
\end{eqnarray}
and using Eq.~(\ref{eq:bog1}), we can write
\begin{eqnarray}
\left\langle c_{j}^{\dagger}c_{j+n}\right\rangle & = &\frac{1}{2N} \sum_{|k|<\frac{\pi}{2}}e^{ikn}\left[(1+(-1)^{n}) + (1- (-1)^{n})\cos\theta_{k} \right.\nonumber\\
&& - \left.(-1)^{j}(1+(-1)^{n})\sin\theta_{k}\right],
\end{eqnarray}
which makes use of the ground state expectation values $\left\langle \Phi_{0}\right|\gamma_{k}^{\dagger}\gamma_{k}\left|\Phi_{0}\right\rangle = 1$, $\left\langle \Phi_{0}\right|\gamma_{k+\pi}^{\dagger}\gamma_{k+\pi}\left|\Phi_{0}\right\rangle = 0$, for $|k|<\frac{\pi}{2}$. We work in the thermodynamic limit $N\rightarrow\infty$ where we may convert the sum over $k$ to an integral and combine terms so that the range may be extended to the entire first Brillouin zone of the gapless initial Hamiltonian, $k\in (-\pi,\pi)$. Noting the equivalent mapping $\left(\pi, \frac{3\pi}{2}\right) \rightarrow \left(-\pi, -\frac{\pi}{2}\right)$, we ultimately obtain
\begin{eqnarray}
\left\langle c_{j}^{\dagger}c_{j+n}\right\rangle & = & \frac{1}{2}\int_{-\pi}^{\pi}\frac{dk}{2\pi}e^{ikn}\left[1+\cos\theta_{k}-(-1)^{j}\sin\theta_{k}\right].\label{eq:cc0}\nonumber\\
\end{eqnarray}
Observables are most compactly written in terms of Majorana fermions $A_{j}=c_{j}^{\dagger}+c_{j}$ and $B_{j} = c_{j}^{\dagger} - c_{j}$ for which the basic contractions reduce to
\begin{eqnarray}
\left\langle A_{j}A_{j+n}\right\rangle & = & \delta_{n=0},\label{eq:aa0}\\
\left\langle B_{j}A_{j+n}\right\rangle & = & \int_{-\pi}^{\pi}\frac{dk}{2\pi}e^{ikn}\frac{\cos k - (-1)^{j}\tilde{m}}{\sqrt{\cos^{2}k + \tilde{m}^{2}}},\label{eq:ba0}
\end{eqnarray}
where $\tilde{m} \equiv m/J$. The magnetization $\left\langle\hat{S}_{j}^{z}\right\rangle = \frac{1}{2}\left\langle\Phi_{0}\right| B_{j}A_{j}\left|\Phi_{0}\right\rangle = \left\langle\Phi_{0}\right| c^{\dagger}_{j}c_{j}\left|\Phi_{0}\right\rangle-\frac{1}{2}$ is staggered, following the profile of the magnetic field
\begin{eqnarray}
\left\langle \hat{S}_{j}^{z}\right\rangle & = & -\frac{\tilde{m}(-1)^{j}}{\pi\sqrt{1+\tilde{m}^{2}}}\mathcal{K}\left(\frac{1}{1+\tilde{m}^{2}}\right),\label{eq:magz0}
\end{eqnarray}
where $\mathcal{K}(k^{2})$ is the complete elliptic integral of the first kind defined by
\begin{eqnarray}
\mathcal{K}(k^{2}) & = & \int_{0}^{\frac{\pi}{2}}\frac{d\theta}{\sqrt{1-k^{2}\sin^{2}\theta}}.
\end{eqnarray}

The spin-spin correlation functions may also be computed to examine quasi-long range order in the ground state. In equilibrium, one has
\begin{eqnarray}
\mathcal{C}_{0}^{zz}(n) & = & \left\langle \hat{S}_{j}^{z}\hat{S}_{j+n}^{z}\right\rangle \nonumber\\
& = & \frac{1}{4}\left\langle \Phi_{0}\right|B_{j}A_{j}B_{j+n}A_{j+n}\left|\Phi_{0}\right\rangle,\label{eq:szsz}\\
\mathcal{C}_{0}^{xx}(n) & = & \left\langle\hat{S}_{j}^{x}\hat{S}_{j+n}^{x}\right\rangle\nonumber\\
 & = & \frac{1}{4}\left\langle \Phi_{0}\right|B_{j}A_{j+1}B_{j+1}\cdots B_{j+n-1}A_{j+n}\left|\Phi_{0}\right\rangle.\label{eq:sxsx}
\end{eqnarray}
In the literature, $\mathcal{C}_{0}^{zz}(n)$ is often referred to as the ``longitudinal'' spin-spin correlation function, while $\mathcal{C}_{0}^{xx}(n)$ is termed the ``transverse'' spin-spin correlation function. Due to rotational symmetry about the $z$-axis, the transverse spin-spin correlation functions are equivalent
\begin{eqnarray}
\left\langle\hat{S}_{j}^{x}\hat{S}_{j+n}^{x}\right\rangle & = & \left\langle\hat{S}_{j}^{y}\hat{S}_{j+n}^{y}\right\rangle,
\end{eqnarray}
and in the remainder of the paper we will only explicitly refer to $\mathcal{C}^{xx}(n)$. Wick's theorem applied to Eq.~(\ref{eq:szsz}) gives
\begin{eqnarray}
\mathcal{C}^{zz}_{0}(n) &=&  \left\langle \hat{S}_{j}^{z}\right\rangle\left\langle \hat{S}_{j+n}^{z}\right\rangle  + \frac{1}{4}\left(\delta_{n=0} - g^{(j)}_{n}g^{(j+n)}_{-n}\right),
\end{eqnarray}
where $g^{(j)}_{n} = \frac{1}{2}\left\langle \Phi_{0}\right|B_{j}A_{j+n}\left|\Phi_{0}\right\rangle$. An asymptotic expansion of Eq.~(\ref{eq:ba0}) in powers of $\frac{1}{n}$ yields vanishing coefficients, consistent with exponential decay. The integral in Eq.~(\ref{eq:ba0}) can be evaluated in terms of special functions as performed in Ref.~\cite{Okamoto1988} for a similar case, yielding,
\begin{eqnarray}
\tilde{\mathcal{C}}^{zz}_{0}(n) & \simeq & \mathcal{D}_{0}(\tilde{m})e^{-\alpha(\tilde{m})n},\label{eq:czz0}
\end{eqnarray}
where $\alpha(\tilde{m})= \cosh^{-1}(1+2\tilde{m}^{2})$ and the amplitude $\mathcal{D}_{0}(\tilde{m})$ does not depend on $n$. We have defined $\tilde{\mathcal{C}}^{zz}(n)$ for $n>0$ as the ``connected'' correlation function in which the product of magnetizations is subtracted, $\mathcal{C}^{zz}(n) = \left\langle\hat{S}^{z}_{j}\right\rangle  \left\langle\hat{S}^{z}_{j+n}\right\rangle  + \tilde{\mathcal{C}}^{zz}(n)$.

\begin{figure}
\begin{center}
\includegraphics[width=8.6cm]{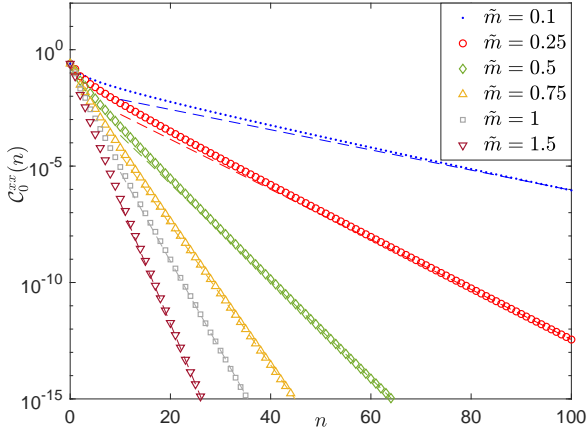}
\caption{Correlation function $\mathcal{C}^{xx}_{0}(n)$ computed in the ground state of Eq.~(\ref{eq:hstag}) for various value of $\tilde{m}$. For $\tilde{m} =0$ (gapless $XY$ model), the correlations decay as a power law $\mathcal{C}_{0}^{xx}(n)\propto n^{-\frac{1}{2}}$, while for $\tilde{m}\neq 0$ the decay is generically exponential with distance $n$. Dashed lines depict exponential decay described by the asymptotic form in Eq.~(\ref{eq:asym1}).}
\label{fig:cxx0stag}
\end{center}
\end{figure}

The transverse spin correlation function given by Eq.~(\ref{eq:sxsx}) involves a string of fermion operators and can be written as a Pfaffian which reduces to a determinant in equilibrium~\cite{Barouch2},
\begin{eqnarray}
\mathcal{C}^{xx}_{0}(n) & = & \frac{1}{4}\left|\begin{array}{cccc} g^{(j)}_{1} & g^{(j)}_{0} & \cdots & g^{(j)}_{-n+2}\\ g^{(j+1)}_{2} & g^{(j+1)}_{1} & \cdots & g^{(j+1)}_{-n+1}\\ \vdots & \vdots & \ddots & \vdots \\
g^{(j+n-1)}_{n} & g^{(j+n-1)}_{n-1} & \cdots & g^{(j+n-1)}_{1}\end{array}\right|.\label{eq:cxxdet}
\end{eqnarray}
In general, Eq.~(\ref{eq:cxxdet}) may be computed numerically. Often, it is possible to apply the Szeg\H{o} limit theorem and related Fisher-Hartwig conjecture to extract the asymptotic exponential or power-law decay analytically for large $n$~\cite{FH1,FH2,FH4}. Such a procedure only applies when ${\bf G}(n)$ is a Toeplitz matrix, having entries $G_{ij}(n) = g_{j-i}$ $(1\leq i,j\leq N)$, which is not the case here due to the term in Eq.~(\ref{eq:ba0}) proportional to $(-1)^{j}$. Such a procedure does work for the $XY$ model~\cite{Ovchinnikov2007}. However, often one may still use the framework of the Fisher-Hartwig conjecture to understand how basic features of the correlations arise. Writing
\begin{eqnarray}
g_{n}^{(j)} & = & \int_{-\pi}^{\pi}\frac{dk}{2\pi}e^{-ikn}\tilde{g}^{(j)}(k),\label{eq:corrft}
\end{eqnarray}
the asymptotic determinant of ${\bf G}$ with entries $G^{(j)}_{il} = g^{(j)}_{l-i}$ and size $n$ can be extracted from the properties of $\tilde{g}^{(j)}(k)$ when the Toeplitz condition is satisfied. In that case, $\tilde{g}(k)$ can often be factored into a product of some smooth part $f_{0}(k)$, and a finite number of zeros and jump discontinuities occurring at points $k_{l}$,
\begin{eqnarray}
\tilde{g}(k) & = & f_{0}(k)\prod_{l}t_{\beta_{l}}(k-k_{l})\left(2-2\cos k_{l}\right)^{\alpha_{l}}.\label{eq:fh1}
\end{eqnarray}
The jump discontinuities are parametrized by numbers $\beta_{l}$ through the function $t_{\beta_{l}}(k-k_{l}) \equiv \exp\left[i\beta_{l}\left(k-k_{l}-\pi\mbox{sgn}(k-k_{l})\right)\right]$. Given the form in Eq.~(\ref{eq:fh1}), the Fisher-Hartwig conjecture states that the asymptotic form of the determinant of ${\bf G}(n)$ is given by
\begin{eqnarray}
\mbox{det}{\bf G}(n) & \sim & \mathcal{M}e^{\overline{f}_{0}n}n^{\sum_{l}(\alpha_{l}^{2}-\beta_{l}^{2})}\;\;\;\;\;\;\;(\mbox{as }n\rightarrow\infty),\label{eq:fhform}
\end{eqnarray}
where $\mathcal{M}$ is a constant independent of $n$ and
\begin{eqnarray}
\overline{f}_{0} & = & \int_{-\pi}^{\pi}\frac{dk}{2\pi}\ln f_{0}(k).
\end{eqnarray}
A procedure exists for calculating $\mathcal{M}$, but it is quite involved whenever $\overline{f}_{0}\neq 0$~\cite{Ovchinnikov2007}. The additional position dependence $\tilde{g}(k) \rightarrow\tilde{g}^{(j)}(k)$ in the present case, where
\begin{eqnarray}
\tilde{g}^{(j)}(k) & = & \frac{\cos k - (-1)^{j}\tilde{m}}{\sqrt{\cos^{2}k + \tilde{m}^{2}}},\label{eq:gf}
\end{eqnarray}
prevents one from applying the Fisher-Hartwig conjecture directly. However, formally factoring Eq.~(\ref{eq:ba0}) into a smooth nonzero function and a product of zeros and jump discontinuities gives
\begin{eqnarray}
\tilde{g}^{(j)}(k) & = & \frac{1}{2\sqrt{\cos^{2}k + \tilde{m}^{2}}}\sqrt{2(1-\cos(k-k_{0}^{(j)}))}\nonumber\\
& & \times\sqrt{2(1-\cos(k+k_{0}^{(j)}))}t_{\frac{1}{2}}(k-k^{(j)}_{0})t_{\frac{1}{2}}(k+k_{0}^{(j)}).\nonumber\\\label{eq:genf}
\end{eqnarray}
where $\cos k_{0}^{(j)} \equiv (-1)^{j}\tilde{m}$, and we take $\tilde{m}<1$. Naively attempting to apply the Fisher-Hartwig conjecture, one finds Eq.~(\ref{eq:genf}) consists of two zeros, each with $\alpha = \frac{1}{2}$ and two jump discontinuities, each with $\beta = \frac{1}{2}$. Applying Eq.~(\ref{eq:fhform}) predicts a vanishing power-law factor. The smooth envelope $f_{0}(k)$ results in a nonzero, exponential decay constant 
\begin{eqnarray}
\overline{f}_{0} & =&  \int_{0}^{2\pi} \frac{dk}{2\pi}\ln \left[\frac{1}{2\sqrt{\cos^{2}k + \tilde{m}^{2}}}\right] = - \sinh^{-1}\tilde{m},
\end{eqnarray}
so that one might expect
\begin{eqnarray}
\mathcal{C}^{xx}_{0}(n) & \sim & \mathcal{A}_{0}(\tilde{m})e^{-n\sinh^{-1}\tilde{m}} \nonumber\\
& = & \mathcal{A}_{0}(\tilde{m})\left(\tilde{m} + \sqrt{\tilde{m}^{2} + 1}\right)^{-n},\;\;\;\;(\mbox{as }n\rightarrow\infty),\label{eq:asym1}
\end{eqnarray}
where $\mathcal{A}_{0}(\tilde{m})$ does not depend on $n$. Remarkably, this simple exponential decay is in agreement with a direct evaluation of Eq.~(\ref{eq:cxxdet}) obtained by integrating Eq.~(\ref{eq:corrft}) numerically using Eq.~(\ref{eq:gf}). A comparison of the numerical evaluation of $\mathcal{C}^{xx}_{0}(n)$ to Eq.~(\ref{eq:asym1}) is shown in Fig.~\ref{fig:cxx0stag}. It is noteworthy that Eq.~(\ref{eq:asym1}) is obtained as a speculative prediction based on the structure of the generating function in Eq.~(\ref{eq:gf}). It should be emphasized that the Fisher-Hartwig conjecture is not applicable to non-Toeplitz matrices, so it is somewhat of a fortunate accident that the correct correlation length $\ell = 1/\sinh^{-1}\tilde{m}$ is obtained by its use here. Similar casual applications of the Fisher-Hartwig conjecture outside its domain of validity will not lead to correct predictions in the next sections. We also note that the correct amplitudes $\mathcal{A}_{0}(\tilde{m})$ do not result from this type of analysis. The values of $\mathcal{A}_{0}(\tilde{m})$ used in Fig.~\ref{fig:cxx0stag} are found by fitting Eq.~(\ref{eq:asym1}) to the numerical evaluation of $\mathcal{C}^{xx}_{0}(n)$, while the decay constant is fully fixed by Eq.~(\ref{eq:asym1}).

The exponential decay observed in $\mathcal{C}_{0}^{xx}(n)$ and $\tilde{\mathcal{C}}_{0}^{zz}(n)$ is an expected feature of a correlation functions in the ground state of a system with a nonzero energy gap. As $\tilde{m}\rightarrow 0$, the energy gap vanishes and both correlation functions reduce to power laws. Note that there is no qualitative change in behavior as $\tilde{m}$ is increased above unity, where $\tilde{m} = 1$ corresponds to $m = J$. Larger values of $\tilde{m}$ lead to difficulties in accurate numerical calculations, as the correlations decay so sharply with $n$. For this reason, we restrict our attention to $\tilde{m} <1$ in the remainder of the paper, as no qualitative differences in behavior have been observed for $\tilde{m}>1$. Away from equilibrium, we will find the structure of the generating function in Eq.~(\ref{eq:gf}) to change in such a way that asymptotic power-law decay can also be roughly observed to emerge in the scenario considered in Sec.~\ref{sec:dynamics}.

\subsection{Dimerized hopping}
An alternative way of generating an energy gap leading to qualitatively similar consequences for physical observables is to consider the dimerized, isotropic $XY$ chain~\cite{Pincus1971,Okamoto1988,Campos2007} in which the nearest-neighbor $XY$ coupling $J$ oscillates in strength, $J\rightarrow J_{j} = J(1-(-1)^{j}\delta)$ so that
\begin{eqnarray}
\hat{H}_{d} & = & -J\sum_{j}\left[\left(1-(-1)^{j}\delta\right)\left(\hat{S}_{j}^{x}\hat{S}_{j+1}^{x} + \hat{S}_{j}^{y}\hat{S}_{j+1}^{y}\right)\right],\\
& = & -\frac{J}{2}\sum_{j}\left[(1-(-1)^{j}\delta)\left(c_{j}^{\dagger}c_{j+1} + c_{j+1}^{\dagger}c_{j}\right)\right].\label{eq:hdim}
\end{eqnarray}
Here $0\leq \delta <1$, and the limit $\delta\rightarrow 1$ results in the system decoupling into isolated pairs of spins. Only values of $\delta$ less than unity will be considered in the remainder of this paper. Equation~(\ref{eq:hdim}) follows from the Jordan-Wigner transformation in Eqs.~(\ref{eq:jw1})--(\ref{eq:jw3}). The Hamiltonian in Eq.~(\ref{eq:hdim}) is diagonalized by the same basic procedure used in the previous section. Defining
\begin{eqnarray}
\left(\begin{array}{c} c_{k}\\ c_{k+\pi}\end{array}\right) & = & \left(\begin{array}{cc} \cos\frac{\phi_{k}}{2} & i\sin\frac{\phi_{k}}{2}\\ i\sin\frac{\phi_{k}}{2} & \cos\frac{\phi_{k}}{2}\end{array}\right)\left(\begin{array}{c} \xi_{k} \\ \xi_{k+\pi}\end{array}\right),\label{eq:bog2}
\end{eqnarray}
we find a diagonal Hamiltonian
\begin{eqnarray}
\hat{H}_{d} & = & -\sum_{|k|<\frac{\pi}{2}}\lambda_{k}\left[\xi_{k}^{\dagger}\xi_{k} - \xi_{k+\pi}^{\dagger}\xi_{k+\pi}\right],
\end{eqnarray}
with $\lambda_{k} = J\sqrt{\cos^{2}k + \delta^{2}\sin^{2}k}$. Diagonalization corresponds to a choice of Bogoliubov angle given by
\begin{eqnarray}
\tan\phi_{k} & = & \delta \tan k,\label{eq:tan2}
\end{eqnarray}
for which $|k|<\frac{\pi}{2}$. As in the previous section, the ground state $\left|\chi_{0}\right\rangle$ contains all negative-energy quasiparticle modes,
\begin{eqnarray}
\left|\chi_{0}\right\rangle & = & \prod_{|k|<\frac{\pi}{2}}\xi_{k}^{\dagger}\left|0\right\rangle.
\end{eqnarray}
Repeating the same basic steps as in the previous section, one finds the basic Majorana contractions are given by
\begin{eqnarray}
\left\langle A_{j}A_{j+n}\right\rangle & = & -\left\langle B_{j}B_{j+n}\right\rangle = \delta_{n=0},\\
\left\langle B_{j}A_{j+n}\right\rangle & = & \int_{-\pi}^{\pi}\frac{dk}{2\pi}e^{ikn}\frac{\cos k +i(-1)^{j}\delta \sin k}{\sqrt{\cos^{2}k + \delta^{2}\sin^{2}k}}.
\end{eqnarray}
With no external magnetic field, one obtains a vanishing magnetization in the ground state
\begin{eqnarray}
\left\langle \hat{S}_{j}^{z}\right\rangle & = & \frac{1}{2}\left\langle \chi_{0}\right|B_{j}A_{j}\left|\chi_{0}\right\rangle = 0.
\end{eqnarray}
The longitudinal correlation function has been evaluated previously~\cite{Okamoto1988} with
\begin{eqnarray}
\mathcal{C}^{zz}_{0}(n) & \sim & \mathcal{D}'_{0}(\delta)e^{-\beta(\delta)n},\label{eq:czz2}
\end{eqnarray}
as $n\rightarrow \infty$ where $\beta(\delta) \equiv \log \frac{1+\delta}{1-\delta}$ and $\mathcal{D}'_{0}(\delta)$ does not depend on $n$. 

Interestingly, attempting to extract the correlation length $\ell$ for the transverse correlation function by appealing to the Fisher-Hartwig conjecture outside its domain of validity as before leads to the incorrect conclusion that $\ell \rightarrow 1/\ln(1+\delta)$. While this {\it does} agree with the numerical evaluation of $\mathcal{C}_{0}^{xx}(n)$ for small $\delta$ and $n\rightarrow \infty$, it fails as $\delta\rightarrow 1$. On general grounds, one expects $\ell \rightarrow 0$ as $\delta\rightarrow 1$ since this limit corresponds to pairs of spins which become decoupled from the rest of the system. As the Fisher-Hartwig conjecture simply does not apply to non-Toeplitz matrices, that it {\it does} work in the case of a staggered magnetic field it is more surprising than that it does not give the correct correlation length in the present setting. 

\begin{figure}
\begin{center}
\includegraphics[width=8.6cm]{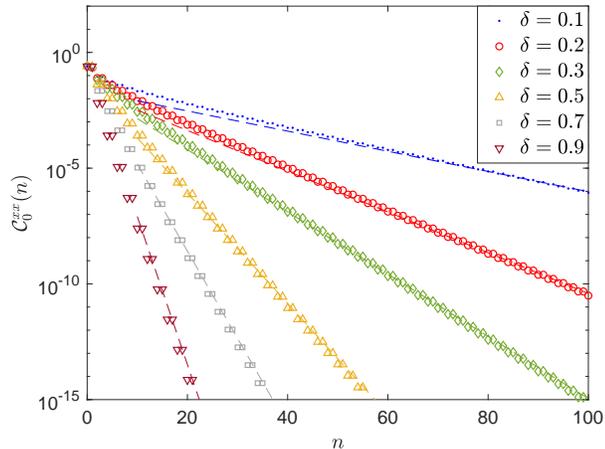}
\caption{Correlation function $\mathcal{C}^{xx}_{0}(n)$ computed in the ground state of Eq.~(\ref{eq:hdim}) for various values of $\delta$. For $\delta\neq 0$ the decay is generically exponential with distance $n$. Dashed lines depict the exponential decay described by the asymptotic form in Eq.~(\ref{eq:cxx2}).}
\label{fig:cxx0dim}
\end{center}
\end{figure}

However, taking exponential decay with exactly {\it half} the correlation length of $\mathcal{C}^{zz}_{0}(n)$ does give an asymptotic form for the transverse correlation function which agrees quite well with a numerical evaluation of $\mathcal{C}^{xx}_{0}(n)$,
\begin{eqnarray}
\mathcal{C}^{xx}_{0}(n) & \sim & \mathcal{A}'_{0}(\delta)\left(\frac{1+\delta}{1-\delta}\right)^{-\frac{n}{2}},\label{eq:cxx2}
\end{eqnarray}
as $n\rightarrow\infty$. The expression in Eq.~(\ref{eq:cxx2}) is shown alongside a numerical evaluation of $\mathcal{C}^{xx}_{0}(n)$ along the same lines as that discussed in the previous section for several values of $\delta$ in Fig.~\ref{fig:cxx0dim}. The correlation function decays exponentially with $n$ for all values of $\delta \neq 0$, and Eq.~(\ref{eq:cxx2}) correctly captures the strength of the exponential decay. Again, the amplitudes $\mathcal{A}'_{0}$ are obtained by fitting Eq.~(\ref{eq:cxx2}) to the numerical evaluation of $\mathcal{C}^{xx}_{0}(n)$. Additionally, the effects of dimerization are seen in the ``staircase'' structure of $\mathcal{C}^{xx}_{0}(n)$ which becomes more prominent as $\delta\rightarrow 1$.

\section{Dynamics from domain-wall initial state}
\label{sec:dynamics}
The main focus of this work concerns how the equilibrium correlation functions presented in the previous section are modified when the systems are far from equilibrium. To drive the system into a nonequilibrium steady state, we assume an initial state $\left|\Psi_{0}\right\rangle$ corresponding to a domain-wall magnetization profile,
\begin{eqnarray}
\left|\Psi_{0}\right\rangle & = & \left|\cdots \uparrow\uparrow\uparrow\downarrow\downarrow\downarrow \cdots\right\rangle\label{eq:dwstate},\\
\left\langle \hat{S}_{j}^{z}\right\rangle & = & \left\{\begin{array}{cc} + \frac{1}{2} & (j\leq 0)\\ -\frac{1}{2} & (j>0)\end{array}\right.
\end{eqnarray}
More general domain walls in which the transition region has nonzero width or the system is only partially polarized far from the central regions have also been considered in the $XY$ model~\cite{Antal1999,Lancaster2010a,Sabetta2013,Lancaster2016}. The former situation has no effect on the long-time dynamics or formation of the steady state. The latter scenario in which the system halves are only partially polarized is a straightforward extension of the main results in this paper. The basic expressions needed to compute observables are somewhat lengthy and relegated to Appendix~\ref{sec:app2}. Restricting attention to the state in Eq.~(\ref{eq:dwstate}), we may write $\left|\Psi_{0}\right\rangle$ in terms of Jordan-Wigner fermions by using the site basis and only occupying sites on the left half of the system,
\begin{eqnarray}
\left|\Psi_{0}\right\rangle & = & \prod_{j\leq 0}c_{j}^{\dagger}\left|0\right\rangle.\label{eq:psi0}
\end{eqnarray}
Our main interest for the remainder of this section is in computing the long-time limit of observables such as spin current and correlation functions with the initial state $\left|\Psi_{0}\right\rangle$ which is not an eigenstate of either Hamiltonian (c.f., Eqs.~(\ref{eq:hstag}),~(\ref{eq:hdim})).

At long times, a non-equilibrium steady state (NESS) forms in the central region of the system in which the magnetization relaxes to zero and the spin current saturates to its asymptotic value. Local observables within this region may be computed in terms of the basic fermionic correlation function
\begin{eqnarray}
\left\langle c_{j}^{\dagger}c_{j+n}\right\rangle_{\mbox{\scriptsize NESS}} & = & \int \frac{dp}{2\pi}e^{-ipn}G(p),
\end{eqnarray}
where $G(p)$ is the {\it effective} momentum distribution of a nonequilibrium steady state which carries information about the initial domain-wall configuration. This framework has been applied previously to study relaxation of observables in the isotropic and anisotropic $XY$ models~\cite{Sabetta2013,Lancaster2016}. Given $\left\langle c_{j}^{\dagger}c_{j+n}\right\rangle_{\mbox{\scriptsize NESS}}$, the long-time limits of all local observables may be computed by application of Wick's theorem. What distinguishes the focus of this work from these previous efforts is the broken translation invariance through the staggered magnetic field or modulated hopping amplitude. These complications result in a position dependence being inherited by the effective momentum distribution, $G(p)\rightarrow G^{(j)}(p)$. The proper interpretation of $G^{(j)}(p)$ is actually the Wigner distribution~\cite{Sabetta2013}, though the distinction is largely unimportant in cases without explicit position dependence.

The overall strategy in this section is to exploit the quadratic nature of each Hamiltonian to obtain explicit expressions for the time-dependent fermion operators $c_{j}(t)$ and compute the basic two-point function $\left\langle c_{j}^{\dagger}(t)c_{j+n}(t)\right\rangle$. As pointed out in Ref.~\onlinecite{Sabetta2013}, the long-time limit will depend on momentum correlations in the initial state, $\left\langle \Psi_{0}\right| c_{k}^{\dagger}c_{k'}\left|\Psi_{0}\right\rangle$, with $k = p+\frac{q}{2}$, $k' = p-\frac{q}{2}$ and $q$ small but nonvanishing. In this limit, 
\begin{eqnarray}
\left\langle \Psi_{0}\right| c_{p+\frac{q}{2}}^{\dagger}c_{p-\frac{q}{2}}\left|\Psi_{0}\right\rangle & \simeq & \frac{-i}{q-i0^{+}}.\label{eq:pole}
\end{eqnarray}
The pole and residue structure encoded in Eq.~(\ref{eq:pole}) contain all the information needed to extract the long-time limit of the Wigner distribution $G^{(j)}(p)$ in the non-equilibrium steady state. As the models considered are quadratic, Wick's theorem reduces all observables to functions of this basic distribution function. The form given in Eq.~(\ref{eq:pole}) is specialized to the fully-polarized semi-infinite subsystems considered here. Appendix~\ref{sec:app2} contains a discussion regarding generalization to domain walls of arbitrary heights. 

\subsection{Staggered magnetic field}
\noindent To compute the Wigner distribution $G^{(j)}(p)$ in the long-time limit, we must first obtain expressions for the time-evolved fermion operators $c_{j}(t)$ in the Heisenberg picture. Given the Hamiltonian Eq.~(\ref{eq:hstag}) and Bogoliubov rotation in Eq.~(\ref{eq:bog1}), the time-evolved position-basis operators can be written as
\begin{eqnarray}
c_{j}(t) & = & \frac{1}{\sqrt{L}}\sum_{|k|<\frac{\pi}{2}}e^{ikj}\left[(f_{kt}+(-1)^{j}g_{kt})c_{k} \right.\nonumber\\
& + & \left. ((-1)^{j}f_{kt}^{*}+g_{kt})c_{k+\pi}\right],
\end{eqnarray}
where
\begin{eqnarray}
f_{kt} & = & \cos(\epsilon_{k}t) - i\cos\theta_{k}\sin(\epsilon_{k}t),\\
g_{kt} & = & i\sin\theta_{k}\sin(\epsilon_{k}t),
\end{eqnarray}
with $\epsilon_{k} = J\sqrt{\cos^{2}k +\tilde{m}^{2}}$. The exact expression for the basic two point function is
\begin{eqnarray}
& & \left\langle c_{j}^{\dagger}(t)c_{j+n}(t)\right\rangle = \int_{-\frac{\pi}{2}}^{\frac{\pi}{2}}\frac{dk}{2\pi}\frac{dk'}{2\pi}e^{-ijk+ik'(j+n)}\left\{\right.\nonumber\\
& & \left[f_{kt}^{*}+(-1)^{j}g_{kt}^{*}\right]\left[f_{k't}+(-1)^{j+n}g_{k't}\right]\left\langle c_{k}^{\dagger}c_{k'}\right\rangle \nonumber\\
& + & \left.\left[(-1)^{j}f_{kt}+g_{kt}^{*}\right]\left[(-1)^{j+n}f_{k't}^{*}+g_{k't}\right]\left\langle c_{k+\pi}^{\dagger}c_{k'+\pi}\right\rangle\right\}.\nonumber\\\label{eq:cjt}
\end{eqnarray}
All expectation values are with respect to the initial state $\left|\Psi_{0}\right\rangle$. Cross terms proportional to $\left\langle c_{k}^{\dagger}c_{k'+\pi}\right\rangle$ and $\left\langle c_{k+\pi}^{\dagger}c_{k'}\right\rangle$ have been dropped, as these contractions give negligible contributions (see Eq.~(\ref{eq:pole})).  Changing variables of integration from $(k,k')$ to $(p,q)$ via $k = p+\frac{q}{2}$ and $k' = p-\frac{q}{2}$, using Eq.~(\ref{eq:pole}) lets us perform the integration over $q$ in the long time limit $t\rightarrow \infty$ as a contour integral, yielding the NESS result
\begin{eqnarray}
\left\langle c_{j}^{\dagger}c_{j+n}\right\rangle_{\mbox{\scriptsize NESS}} & = & \lim_{t\rightarrow\infty}\left\langle c_{j}^{\dagger}(t)c_{j+n}(t)\right\rangle\\
& = & \int_{-\pi}^{\pi}\frac{dp}{2\pi}e^{-ipn}G^{(j)}(p),\label{eq:nessck0}
\end{eqnarray}
where the Wigner function $G^{(j)}(p)$ is given by
\begin{eqnarray}
G^{(j)}(p) & = & \frac{1}{2}\left[1 +\left(\cos\theta_{p}-(-1)^{j}\sin\theta_{p}\right)\sigma(p)\right],\label{eq:nessck1}\\
\sigma(p) & = & \mbox{sgn}(p)\mbox{sgn}\left(\frac{\pi}{2}-p\right)\mbox{sgn}\left(\frac{\pi}{2}+p\right).\label{eq:nessck2}
\end{eqnarray}
Equation~(\ref{eq:nessck1}) underlies all of the remaining results in this section, and its form is strikingly similar to the equilibrium result (c.f. Eq.~(\ref{eq:cc0})). In taking the formal limit $t\rightarrow \infty$, our results apply at large but finite times for $|j|, |j+n| \ll Jt$. Indeed, $G^{(j)}(p)$ relaxes in the long-time limit to its equilibrium form but with the additional factor $\sigma(p)$ which is equal to $\pm 1$ and consists of four piecewise-constant segments. This combination of step functions arises in part from projecting the integral to the entire range $-\pi <p<\pi$. This series of sign changes provided by $\sigma(p)$ has dramatic consequences for the observables and contains all of the surviving information about the initial state.

\subsubsection{Magnetization and spin current}
From Eq.~(\ref{eq:cjt}) one may compute observables at arbitrary times. Given that $\hat{H}_{s}$ maps to free fermions in Eq.~(\ref{eq:hstag1}) and that the initial state in Eq.~(\ref{eq:psi0}) can be represented by the ground state of a different quadratic Hamiltonian, it is possible to obtain the corresponding results for a finite system from a sequence of direct matrix diagonalizations, as described in Appendix~\ref{sec:app1}. Figure~\ref{fig:j1} shows the magnetization dynamics for various choices of $\frac{m}{J}$ with a sharp domain-wall initial state. The transient dynamics obtained numerically from the method described in Appendix~\ref{sec:app1} are virtually indistinguishable from the analytic formula given in Eq.~(\ref{eq:cjt}).
\begin{figure}
\begin{center}
\includegraphics[width=8.6cm]{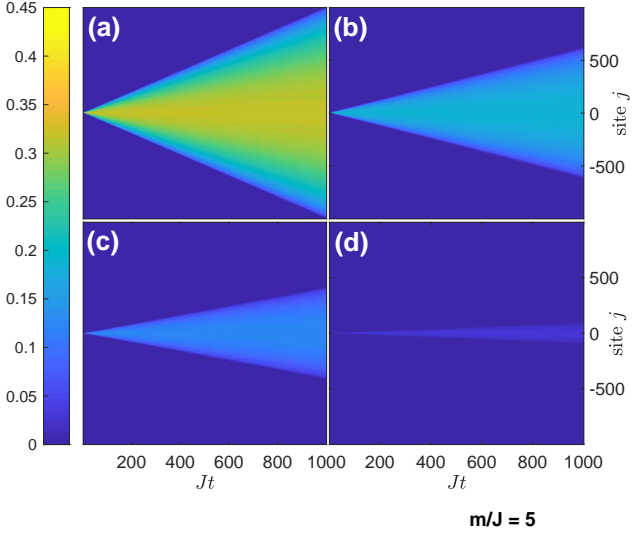}
\caption{Time evolution of spin current for different values of $\tilde{m} \equiv m/J$ with (a) $\tilde{m} = 0$, (b) $\tilde{m}=\frac{1}{2}$, (c) $\tilde{m} = 1$, and (d) $\tilde{m} = 5$. The ``light cone'' swept by the nonzero current corresponds to the growing central region in which the magnetization relaxes (asymptotically) to zero. As $m/J$ becomes larger, the domain wall spreads more slowly and the resulting steady-state current becomes smaller.}
\label{fig:j1}
\end{center}
\end{figure}
Using Eq.~(\ref{eq:nessck1}), the long-time limit of the magnetization in the NESS is shown to vanish,
\begin{eqnarray}
\left\langle \hat{S}_{j}^{z}\right\rangle_{\mbox{\scriptsize NESS}} & = & \left\langle c_{j}^{\dagger}c_{j}\right\rangle_{\mbox{\scriptsize NESS}} - \frac{1}{2}\\
& = & 0.
\end{eqnarray}
In the case of $m=0$, this vanishing magnetization represents relaxation to the ground state average in the isotropic $XY$ spin chain. In the gapped model, the ground state possesses a staggered magnetization given by Eq.~(\ref{eq:magz0}). Thus, while the magnetization relaxes to zero, the vanishing magnetization is quite different from the equilibrium profile, showing that the system remains in a highly excited state of the Hamiltonian $\hat{H}_{s}$. 

The spin current $\hat{\mathcal{J}}_{j}^{z}$, which vanishes in the system's ground state, becomes nonzero due to the net flow of magnetization from the left side to the right side of the system. The precise form of $\hat{\mathcal{J}}^{z}_{j}$ follows from the continuity equation for the magnetization,
\begin{eqnarray}
\partial_{t}\hat{S}_{j}^{z} & = & -\left(\hat{\mathcal{J}}^{z}_{j+1}-\hat{\mathcal{J}}_{j}^{z}\right).
\end{eqnarray}
Applying the Heisenberg equation of motion for time evolution generated by the Hamiltonian $\hat{H}_{s}$ in Eq.~(\ref{eq:hstag}),
\begin{eqnarray}
\partial_{t}\hat{S}_{j}^{z} & = & i\left[\hat{H}_{s},\hat{S}_{j}^{z}\right],\label{eq:heiseom}
\end{eqnarray}
one can identify
\begin{eqnarray}
\hat{\mathcal{J}}_{j}^{z} & = & J\left[\hat{S}_{j+1}^{+}\hat{S}_{j}^{-} - \hat{S}_{j}^{+}\hat{S}_{j+1}^{-}\right]\\
& = & \frac{iJ}{2}\left[c_{j+1}^{\dagger}c_{j} - c_{j}^{\dagger}c_{j+1}\right],\label{eq:current}
\end{eqnarray}
which is the same as the expression for spin current in the isotropic $XY$ model~\cite{Lancaster2016}. For a domain-wall initial state, Eq.~(\ref{eq:nessck1}) can be used in Eq.~(\ref{eq:current}) to obtain
\begin{eqnarray}
\left\langle \hat{\mathcal{J}}_{j}^{z}\right\rangle_{\mbox{\scriptsize NESS}} & = & \frac{J}{2i}\int_{-\pi}^{\pi}\frac{dp}{2\pi}\left(e^{ip}-e^{-ip}\right) G(p),\\
& = & \frac{J}{\pi}\left(\sqrt{\tilde{m}^{2}+1}- \tilde{m}\right).\label{eq:curt}
\end{eqnarray}
It should be noted that the algebraic decay in Eq.~(\ref{eq:curt}) is qualitatively similar to that obtained from a quench from an initial state which is spatially homogeneous but supports a nonzero spin current~\cite{Lancaster2010b}.

\subsubsection{Correlations}
To obtain the general equal-time spin correlation functions, it is convenient to work in terms of the Majorana contractions, which can be written as
\begin{eqnarray}
\left\langle B_{j}A_{j+n}\right\rangle & = & \left\langle c_{j}^{\dagger}c_{j+n}\right\rangle + \left\langle c_{j+n}^{\dagger}c_{j}\right\rangle -\delta_{n=0},\\
\left\langle A_{j}A_{j+n}\right\rangle & = & -\left\langle B_{j}B_{j+n}\right\rangle  \nonumber\\
& = & \left\langle c_{j}^{\dagger}c_{j+n}\right\rangle - \left\langle c_{j+n}^{\dagger}c_{j}\right\rangle +\delta_{n=0},
\end{eqnarray}
The general time-dependent spin-spin correlation functions now take the form
\begin{eqnarray}
\mathcal{C}^{zz}(n,t) & = & \frac{1}{4}\left\langle \Psi_{0}\right|B_{j}(t)A_{j}(t)B_{j+n}(t)A_{j+n}(t)\left|\Psi_{0}\right\rangle,\label{eq:szszness1}\\
\mathcal{C}^{xx}(n,t) & = & \frac{1}{4}\left\langle \Psi_{0}\right|B_{j}(t)A_{j+1}(t)B_{j+1}(t)\cdots A_{j+n}(t)\left|\Psi_{0}\right\rangle.\label{eq:sxsxness1}
\end{eqnarray}
Our interest here is in the long-time limit after a homogeneous nonequilibrium steady state has formed in the central part of the system where $\mathcal{C}^{\alpha\alpha}(n,t)\rightarrow \mathcal{C}^{\alpha\alpha}_{\mbox{\scriptsize NESS}}(n)$ for $\alpha = x,z$. Using Eqs.~(\ref{eq:nessck0})--(\ref{eq:nessck2}), we obtain
\begin{eqnarray}
 \left\langle B_{j}A_{j+n}\right\rangle_{\mbox{\scriptsize NESS}} & = & 0,\label{eq:ba}\\
\left\langle A_{j}A_{j+n}\right\rangle_{\mbox{\scriptsize NESS}} & = & \int_{-\pi}^{\pi}\frac{dp}{2\pi}e^{-ipn}\nonumber\\
& & \times\left[\frac{\cos p -(-1)^{j}\tilde{m}}{\sqrt{\cos^{2}p + \tilde{m}^{2}}}\right]\sigma(p),\label{eq:aa}\;\;\;\;\;\;\;
\end{eqnarray}
with $\left\langle B_{j}B_{j+n}\right\rangle_{\mbox{\scriptsize NESS}} = -\left\langle A_{j}A_{j+n}\right\rangle_{\mbox{\scriptsize NESS}}$ and $\sigma(p)$ given in Eq.~(\ref{eq:nessck2}). For $n>0$, the longitudinal correlation function becomes
\begin{eqnarray}
\mathcal{C}^{zz}_{\mbox{\scriptsize NESS}}(n) & = & \frac{1}{4}\left[\left\langle A_{j}A_{j+n}\right\rangle_{\mbox{\scriptsize NESS}}\right]^{2}.
\end{eqnarray}
Upon expanding Eq.~(\ref{eq:aa}) for large $n$, we find
\begin{eqnarray}
\mathcal{C}^{zz}_{\mbox{\scriptsize NESS}}(n) & \sim & -\frac{1}{\left(\pi n\right)^{2}}\times \left\{\begin{array}{cc} \displaystyle \frac{1}{1+\tilde{m}^{2}} & (n\mbox{ odd})\\ \displaystyle\left(1- \frac{(-1)^{\frac{n}{2}} \tilde{m}}{\sqrt{1+\tilde{m}^{2}}}\right)^{2} & (n\mbox{ even}),\end{array}\right.\label{eq:czzness}
\end{eqnarray}
with the above expressions holding in the limit $n\rightarrow \infty$. For the domain wall initial state, we observe the persistence of power-law correlations in contrast to the exponentially decaying correlation function in the ground state of the model. Additionally, there are oscillations at multiple wavelengths present in Eq.~(\ref{eq:czzness}), as seen from the alternating functional forms for even and odd $n$ and the factor of $(-1)^{\frac{n}{2}}$ in Eq.~(\ref{eq:czzness}). These intricate oscillations and power-law correlations turn out to be a common feature in the models considered.

Next, we wish to evaluate the transverse correlations. Due to the vanishing of Eq.~(\ref{eq:ba}), the Pfaffian for the transverse correlation function simplifies to
\begin{eqnarray}
\mathcal{C}^{xx}_{\mbox{\scriptsize NESS}}(n)
& = & \frac{1}{4}\mbox{det}{\bf Q},\label{eq:det}
\end{eqnarray}
where ${\bf Q}$ is an antisymmetric matrix with entries given by $Q_{jl} = q_{l-j}^{(j)}$ for $l>j$ with
\begin{eqnarray}
q_{n}^{(j)} & = & \left\langle A_{j}A_{j+n}\right\rangle_{\mbox{\scriptsize NESS}} \label{eq:qmn1}\\
& = & \int_{-\pi}^{\pi}\frac{dp}{2\pi}e^{-ipn}\left[\frac{\cos p -(-1)^{j}\tilde{m}}{\sqrt{\cos^{2}p + \tilde{m}^{2}}}\right]\nonumber\\
& & \times \mbox{sgn}\left(p\right)\mbox{sgn}\left(\frac{\pi}{2}-p\right)\mbox{sgn}\left(\frac{\pi}{2}+p\right).\label{eq:qmn2}
\end{eqnarray}
Antisymmetry defines the entries along and below the diagonal in Eq.~(\ref{eq:qmn1}) while Eq.~(\ref{eq:qmn2}) is valid for all choices of $l$ and $j$. If not for the term proportional to $(-1)^{j}$, ${\bf Q}$ would be a Toeplitz matrix for which the Szeg\H{o} limit theorems and Fisher-Hartwig conjecture~\cite{FH1,FH2,FH4,Ovchinnikov2007} could be used to extract asymptotic behavior of the determinant in Eq.~(\ref{eq:det}) as $|l-j|\rightarrow \infty$. As in equilibrium, it is possible to formally define a generating function for the non-Toeplitz determinant,
\begin{eqnarray}
\tilde{q}^{(j)}(p) & = & \frac{\cos p - (-1)^{j}\tilde{m}}{\sqrt{\cos^{2}p+ \tilde{m}^{2}}}\sigma(p).\label{eq:qjk}
\end{eqnarray}
Comparing Eq.~(\ref{eq:qjk}) to Eq.~(\ref{eq:genf}), the only structural difference between $q^{(j)}_{n}$ and $g^{(j)}_{n}$ is the presence of $\sigma(p)$. This factor of step functions serves to introduce additional jump discontinuities into the generating function. 
One observes that a possible representation for $\sigma(p)$ is
\begin{eqnarray}
\sigma(k) & = & \mbox{sgn}\left(p\right)\mbox{sgn}\left(\frac{\pi}{2}-p\right)\mbox{sgn}\left(\frac{\pi}{2}+p\right),\\
& \propto & t_{\frac{1}{2}}(p)t_{-\frac{1}{2}}(p-\pi)t_{\frac{1}{2}}\left(p-\frac{\pi}{2}\right)t_{-\frac{1}{2}}\left(p+\frac{\pi}{2}\right),
\end{eqnarray}
where $t_{\beta}(k-k_{l}) \equiv \exp\left[\i\beta\left(k-k_{l}-\pi\mbox{sgn}(k-k_{l})\right)\right]$. The generating function $\tilde{q}^{(j)}(k)$ should be periodic, and the role of $t_{-\frac{1}{2}}(p-\pi)$ is to encode the jump discontinuity from $p = \pi-0^{+}$ to $p = \pi + 0^{+}\rightarrow -\pi +0^{+}$. Additionally, this factor cancels a lingering factor of $e^{i\frac{\pi}{2}}$ arising from $t_{\frac{1}{2}}(p)$. 

We stress that the Fisher-Hartwig conjecture strictly does not apply due to the $j$ dependence present in Eq.~(\ref{eq:qjk}). However, the exponential decay constant in equilibrium was predicted by its naive application and we proceed here using the Fisher-Hartwig conjecture as a guide to explore how the nonequilibrium correlations might be expected to differ from equilibrium. Using Eq.~(\ref{eq:fh1}), one would expect an additional contribution to the power law exponent of $4\left(-\frac{1}{2^{2}}\right) = -1$ in this nonequilibrium steady state. Recall that the power-law factor was neither predicted by the Fisher-Hartwig conjecture nor actually present in equilibrium (c.f., Eq.~(\ref{eq:asym1})). As before, there is also a nonzero exponential decay factor. Comparing this power-law prediction to numerical evaluation of Eq.~(\ref{eq:det}), one finds that the exponential decay observed in equilibrium vanishes entirely, leaving only algebraic decay predicted by the discontinuities in $\sigma(p)$,
\begin{eqnarray}
\left\langle \hat{S}^{x}_{j}\hat{S}^{x}_{j+n}\right\rangle_{\mbox{\scriptsize NESS}} & \sim & \left[\mathcal{A}(m)+\mathcal{B}(m)\cos\left(\frac{\pi n}{4}\right)\right]\cos\left(\frac{\pi n}{2}\right)\frac{1}{n},\nonumber\\\label{eq:cxxness}
\end{eqnarray}
as $n\rightarrow\infty$. Note that while the Fisher-Hartwig conjecture {\it does} predict the $n^{-1}$ power law in Eq.~(\ref{eq:cxxness}), it also erroneously produces an exponential decay factor which is not observed in the actual correlation function. For that reason, Eq.~(\ref{eq:cxxness}) is effectively an ansatz which agrees well with the actual correlation function and contains the power-law decay predicted by the Fisher-Hartwig conjecture. The oscillating prefactors arise by comparison to numerical evaluation of $\mathcal{C}^{xx}_{\mbox{\scriptsize NESS}}(n)$ discussed below. It is noteworthy that the Fisher-Hartwig conjecture is often unable to predict such oscillations even when the matrix in question is a legitimate Toeplitz matrix~\cite{Lancaster2016}.

\begin{figure}
\begin{center}
\includegraphics[width=8.6cm]{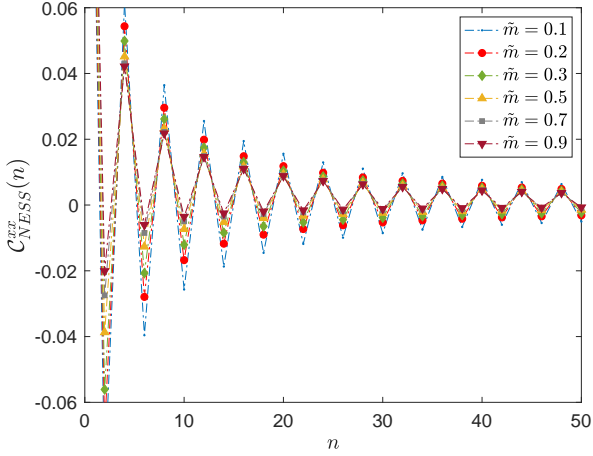}
\caption{ Correlation function $\mathcal{C}^{xx}(n)$ computed in the nonequilibrium steady state of Eq.~(\ref{eq:hdim}) for various values of $\tilde{m}$. Only values for even $n$ are shown, as the correlations vanish for odd $n$.}
\label{fig:cxxosc}
\end{center}
\end{figure}

\begin{figure}
\begin{center}
\includegraphics[width=8.6cm]{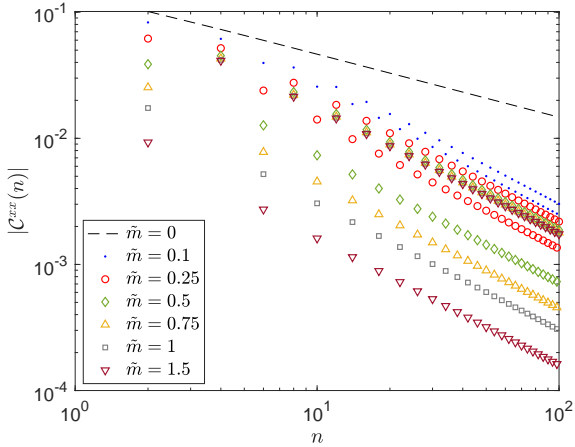}
\caption{Absolute value of correlation function $\mathcal{C}^{xx}(n)$ for even $n$ computed in the nonequilibrium steady state of Eq.~(\ref{eq:hdim}) for various values of $\tilde{m}$.}
\label{fig:cxxnessstag}
\end{center}
\end{figure}

Numerical evaluation of the correlation function for even $n$ is shown in Fig.~\ref{fig:cxxosc} for several values of $\tilde{m}$. Figure~\ref{fig:cxxnessstag} shows the absolute value of $\mathcal{C}^{xx}_{\mbox{\scriptsize NESS}}(n)$, which makes clear the separate branches and power-law decay. The result is that the correlations vanish for all odd $n$, while those for $n$ divisible by 4 decay with a different $m$-dependent amplitude than those with $n$ not divisible by 4. Interestingly, this set of two different amplitudes also appears in a closely-related model in which the effective energy gap is not suddenly switched to its final value but allowed to increase self-consistently as a staggered field is turned on~\cite{Rieger2018,Rieger2018b}. In that work, the initial state was spatially homogeneous so that no persistent current developed at long times. Consequently, the correlation function $\mathcal{C}_{\mbox{\scriptsize NESS}}^{xx}(n)$ showed no oscillations of the order $\cos(\pi n/2)$. However, the even and odd correlations split into separate branches in a manner very similar to Eq.~(\ref{eq:cxxness}). Interestingly, the decay was also algebraic but asymptotically the same as the equilibrium $m=0$ result, with $\mathcal{C}^{xx}(n) \propto n^{-\frac{1}{2}}$.

The main results in this section are the persistence of power-law correlations (c.f., Eqs.~(\ref{eq:czzness}), (\ref{eq:cxxness})) in a nonequilibrium steady state despite the system having exponentially decaying correlations in its ground state. In similar spin systems without the presence of an energy gap, the Fisher-Hartwig conjecture has been applied to extract exact asymptotics of the transverse correlations when beginning from the domain-wall state $\left|\Psi_{0}\right\rangle$~\cite{Lancaster2016}. The presence of a staggered magnetic field leads to the effective momentum distribution in the NESS acquiring a position dependence and assuming the form of a Wigner distribution. This position dependence spoils the Toeplitz nature of the matrix whose determinant gives the transverse correlation function, and the Fisher-Hartwig conjecture can no longer be applied directly. However, by observing how the Wigner distribution $G^{(j)}(p)$ is modified compared to its equilibrium form, one can predict the emergence of an additional power-law decay factor in the NESS correlation function. Interestingly, the exponential decay disappears entirely, leaving an enhanced power-law decay--a fact not easily seen from the explicit form of the Wigner distribution. We shall see below that the dimerized chain leads to extremely similar behavior.

\subsection{Dimerized hopping}
The calculations for the dimerized chain are quite similar to those for the chain with the staggered magnetic field, so we only sketch the main results and elaborate on the new features in this model. Using the Bogoliubov rotation in Eq.~(\ref{eq:bog2}), the time evolution generated by Eq.~(\ref{eq:hdim}) leads to time-dependent position-basis operators
\begin{eqnarray}
c_{j}(t) & = & \frac{1}{\sqrt{L}}\sum_{|k|<\frac{\pi}{2}}e^{ikj}\left[(f_{k}(t)+(-1)^{j}g_{k}(t))c_{k} \right.\nonumber\\
& & \left.+ ((-1)^{j}f_{k}^{*}(t)-g_{k}(t))c_{k+\pi}\right],\label{eq:cjtdim}
\end{eqnarray}
where 
\begin{eqnarray}
f_{kt} & = & \cos(\lambda_{k}t)-i\cos\phi_{k}\sin(\lambda_{k}t),\\
g_{kt} & = & \sin\phi_{k}\sin(\lambda_{k}t).
\end{eqnarray}
Repeating the steps from the previous section, one finds the emergence of a well-defined nonequilibrium steady state,
\begin{eqnarray}
\left\langle c_{j}^{\dagger}c_{j+n}\right\rangle_{\mbox{\scriptsize NESS}} & = & \lim_{t\rightarrow\infty}\left\langle c_{j}^{\dagger}(t)c_{j+n}(t)\right\rangle,\\
& = &  \int_{-\pi}^{\pi}\frac{dp}{2\pi}e^{-ipn}G_{\delta}^{(j)}(p),
\end{eqnarray}
where
\begin{eqnarray}
G_{\delta}^{(j)}(p) & = & \frac{1}{2}\left[1 + \left(\cos\phi_{p}-i(-1)^{j}\sin\phi_{p}\right)\sigma(p)\right],\label{eq:gkdim}
\end{eqnarray}
where $\sigma(p)$ is defined in Eq.~(\ref{eq:nessck2}). Similarly to the previous model considered, the only difference between the nonequilibrium result and the equilibrium form is the presence of $\sigma(p)$, which introduces several jump discontinuities into the Wigner distribution. 
\subsubsection{Magnetization and spin current}
Using Eq.~(\ref{eq:gkdim}), the NESS magnetization is calculated to vanish,
\begin{eqnarray}
\left\langle \hat{S}_{j}^{z}\right\rangle_{\mbox{\scriptsize NESS}} & = & 0,
\end{eqnarray}
which happens to be the equilibrium value for this model. 

One must resort to Eq.~(\ref{eq:heiseom}) to obtain the correct form for the current operator $\hat{\mathcal{J}}^{z}_{j}$ in the presence of nonzero dimerization, obtaining
\begin{eqnarray}
\hat{\mathcal{J}}_{j}^{z} & = & \frac{iJ}{2}\left(1-(-1)^{j}\delta\right)\left[c_{j+1}^{\dagger}c_{j} - c_{j}^{\dagger}c_{j+1}\right].
\end{eqnarray}
The modulating amplitude is compensated by oscillations within the expectation values of the fermion operators, and one finds
\begin{eqnarray}
\left\langle \hat{\mathcal{J}}_{j}^{z}\right\rangle_{\mbox{\scriptsize NESS}} & = & \frac{J(1-\delta)}{\pi}.
\end{eqnarray}
The isotropic $XY$ result is recovered in the limit $\delta\rightarrow 0$. Furthermore, the current vanishes as $\delta \rightarrow 1$, consistent with complete dimerization in which individual pairs become isolated and cease interacting with the rest of the system. 
\subsubsection{Correlations}
\begin{figure}
\begin{center}
\includegraphics[width=8.6cm]{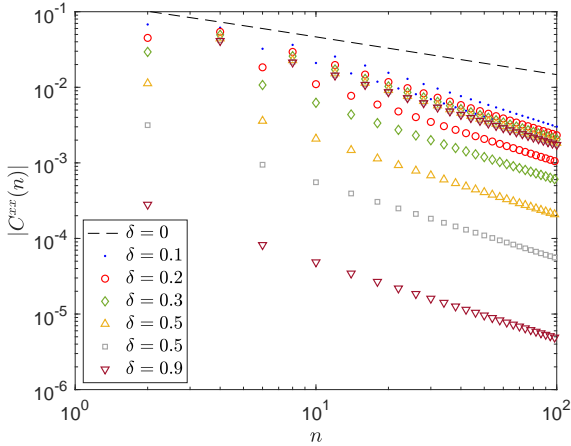}
\caption{Absolute value of correlation function $\mathcal{C}^{xx}(n)$ computed in the nonequilibrium steady state of Eq.~(\ref{eq:hdim}) for various values of $\delta$. The correlations are zero for odd $n$, and these values are omitted from the plot.}
\label{fig:cxxnessdim}
\end{center}
\end{figure}

As in the previous section, all observables follow from Eq.~(\ref{eq:gkdim}). The $\mathcal{C}^{zz}_{\mbox{\scriptsize NESS}}(n)$ correlation function is again the most straightforward to calculate, giving formally the same result as obtained for the staggered field,
\begin{eqnarray}
\mathcal{C}^{zz}_{\mbox{\scriptsize NESS}}(n) & = & \frac{1}{4}\left[\left\langle A_{j}A_{j+n}\right\rangle_{\mbox{\scriptsize NESS}}\right]^{2}.
\end{eqnarray}
For $n\rightarrow\infty$, the asymptotic result is
\begin{eqnarray}
\mathcal{C}^{zz}_{\mbox{\scriptsize NESS}}(n) & \sim & \left\{\begin{array}{cc} 0 & (n\mbox{ even})\\ \displaystyle\left(\frac{2}{\pi n}\right)^{2} & \left(\frac{n-1}{2}\mbox{ even}\right)\\
\displaystyle\left(\frac{1}{\delta}-\delta\right)^{2}\left(\frac{1}{\pi n^{2}}\right)^{2} & \left(\frac{n-1}{2}\mbox{ odd}\right),\end{array}\right.\label{eq:czzness2}
\end{eqnarray}
Again we find power-law decay with oscillations of several wavelengths. Turning attention to $\mathcal{C}^{xx}_{\mbox{\scriptsize NESS}}(n)$, the same factor of $\sigma(p)$ which appeared in the generating function when the staggered magnetic field was switched on speculatively suggests an additional power-law decay factor of $n^{-1}$. Figure~\ref{fig:cxxnessdim} shows the result of numerically evaluating the appropriate determinant for various values of $\delta$, demonstrating that in the nonequilibrium steady state, the exponential decay in the ground state is replaced again by purely algebraic decay of the form $n^{-1}$. As in the previous case, the odd correlations vanish and the even correlations split into two branches with different amplitudes for $\frac{n}{2}$ even and $\frac{n}{2}$ odd,
\begin{eqnarray}
\mathcal{C}^{xx}_{\mbox{\scriptsize NESS}}(n) & \sim & \left[\mathcal{A}'(\delta)+\mathcal{B}'(\delta)\cos\left(\frac{\pi n}{4}\right)\right]\cos\left(\frac{\pi n}{2}\right)\frac{1}{n},\nonumber\\\label{eq:cxxness2}
\end{eqnarray}

To summarize the results for the dimerized chain, we find correlations which are virtually identical to those found in the $XY$ chain with staggered magnetic field. The distinguishing characteristic of both models is an energy gap which opens at $k = \pm \frac{\pi}{2}$, resulting in a doubling of the unit cell and corresponding reduction in the first Brillouin zone. It appears that this reduction in size of the Brillouin zone, which is mathematically the root cause of an enhanced power-law, is intimately related to the stronger correlations appearing in the NESS compared to the ground state. 


\subsection{Quench from ground state of $XY$ model}

\begin{figure}
\begin{center}
\includegraphics[width=8.6cm]{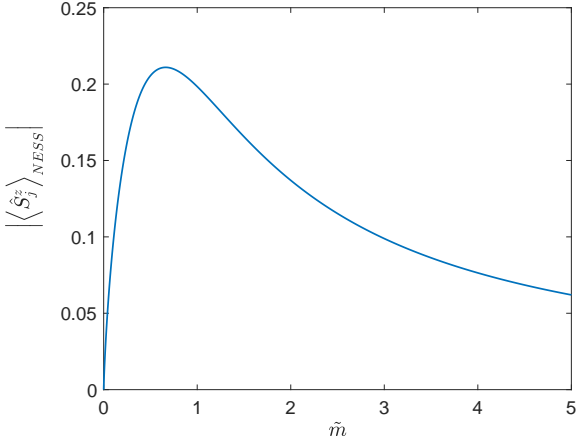}
\caption{Magnitude of staggered magnetization as function of $\tilde{m}$ in non-equilibrium steady state when initial state is the (spatially homogeneous) ground state of the isotropic $XY$ model.}
\label{fig:magzhom}
\end{center}
\end{figure}

The results in the previous session warrant some further investigation into the models considered to understand how generic the power-law correlations are after a quench into gapped phases. Here we explore quenches beginning from the spatially homogeneous ground state of the isotropic $XY$ model without a magnetization domain wall. Again, we find similar behavior when either a staggered magnetic field or dimerization term is switched on, but interestingly the power-law correlations do not emerge. Except where explicitly indicated, similar results are obtained for both the staggered magnetic field and dimerized model. Consequently, results are presented only for the staggered magnetic field. Let us consider the ground state of the $XY$ model, which is obtained from Eq.~(\ref{eq:hstag}) by setting $m = 0$. In this case, the model is diagonalized by simple Fourier transform,
\begin{eqnarray}
\hat{H}_{0} & = & -\frac{J}{2}\sum_{j}\left[c_{j}^{\dagger}c_{j+1}+c_{j+1}^{\dagger}c_{j}\right],\\
& = &- \sum_{k}J\cos k c_{k}^{\dagger}c_{k},
\end{eqnarray}
so that the ground state $\left|\varphi_{0}\right\rangle$ is composed of all negative-energy states,
\begin{eqnarray}
\left|\varphi_{0}\right\rangle & = & \prod_{|k|<\frac{\pi}{2}}c_{k}^{\dagger}\left|0\right\rangle.
\end{eqnarray}
Beginning with the system in the state $\left|\varphi_{0}\right\rangle$, time evolution takes place under $\hat{H}_{s}$ given by Eq.~(\ref{eq:hstag}). Starting from Eq.~(\ref{eq:cjt}) and using
\begin{eqnarray}
\left\langle \varphi_{0}\right| c_{k}^{\dagger}c_{k'}\left|\varphi_{0}\right\rangle & = & \delta_{kk'}\Theta\left(\frac{\pi}{2}-|k|\right),\\
\left\langle \varphi_{0}\right| c_{k+\pi}^{\dagger}c_{k'+\pi}\left|\varphi_{0}\right\rangle & = & 0,
\end{eqnarray}
we find
\begin{eqnarray}
& &\left\langle c_{j}^{\dagger}(t)c_{j+n}(t)\right\rangle = \int_{-\frac{\pi}{2}}^{\frac{\pi}{2}}\frac{dk}{2\pi}e^{ikn}\left[|f_{kt}|^{2} \right.\nonumber\\
& & +\left. (-1)^{n}|g_{kt}|^{2} + (-1)^{j}\left[f_{kt}g_{kt}^{*}+(-1)^{n}f_{kt}^{*}g_{kt}\right]\right].
\end{eqnarray}
The long-time limit can be taken directly, and the terms can be arranged into a single integral over $k$ from $k=-\pi$ to $k=\pi$. Alternatively, one may begin from Eq.~(\ref{eq:wigfinal}) and take the limit $k_{F}^{+} = k_{F}^{-} = \frac{\pi}{2}$. In either case, we find
\begin{eqnarray}
\lim_{t\rightarrow\infty}\left\langle c_{j}^{\dagger}(t)c_{j+n}(t)\right\rangle & = & \left\langle c_{j}^{\dagger}c_{j+n}\right\rangle_{\mbox{\scriptsize NESS}}\\
& = & \int_{-\pi}^{\pi} \frac{dp}{2\pi}e^{-ipn}G^{(j)}_{\mbox{\scriptsize hom.}}(p),
\end{eqnarray}
where
\begin{eqnarray}
G_{\mbox{\scriptsize hom.}}^{(j)}(p) & = & \frac{1}{2} + \frac{1}{2}\left(\cos\theta_{p}-(-1)^{j}\sin\theta_{p}\right)|\cos\theta_{p}|\nonumber\\ \\
& = & \frac{1}{2} + \frac{1}{2}\frac{\cos k -(-1)^{j}\tilde{m}}{\sqrt{\cos^{2}k + \tilde{m}^{2}}}|\cos k|.\label{eq:ghomk}
\end{eqnarray}
The magnetization at long times follows from $\left\langle \hat{S}_{j}^{z}\right\rangle = \frac{1}{2}\left\langle B_{j}A_{j}\right\rangle$ and can be computed exactly,
\begin{eqnarray}
\left\langle \hat{S}_{j}^{z}\right\rangle_{\mbox{\scriptsize NESS}} & = & -\frac{(-1)^{j}\tilde{m}}{\pi \sqrt{1+\tilde{m}^{2}}}\log \left[\frac{\sqrt{1+\tilde{m}^{2}}+1}{\tilde{m}}\right],
\end{eqnarray}
which exhibits a staggered pattern with an amplitude that is not monotonic in $\tilde{m}$, as shown in Fig.~\ref{fig:magzhom}. The peak amplitude occurs for $\tilde{m} = \tilde{m}_{*}$, where $\tilde{m}_{*}\approx 0.6627$ satisfies
\begin{eqnarray}
\log\left[\frac{\sqrt{1+\tilde{m}^{2}}+1}{\tilde{m}}\right] & = & \sqrt{1+\tilde{m}^{2}}.
\end{eqnarray}
Similar such non-monotonic behavior of observables has been found after gapless-to-gapped quenches in a variety of models~\cite{Porta2018}, with a careful investigation of the evolution of correlation functions suggesting that this robust behavior arises from ``freezing'' of the light cone in these systems. 

In the absence of a particle density imbalance provided by the domain-wall initial state, the current vanishes . However, the correlation functions are more interesting. By expanding Eq.~(\ref{eq:ghomk}) in powers of $\frac{1}{n}$, one finds the asymptotic behavior of the correlation function for large $n\rightarrow\infty$ given by 
\begin{eqnarray}
\tilde{\mathcal{C}}^{zz}_{\mbox{\scriptsize NESS}}(n) & \sim  &   \left\{\begin{array}{cc}\displaystyle \frac{1}{\pi^{2}\tilde{m}^{2}n^{4}} & (n\mbox{ even})\\ \displaystyle\frac{4}{\pi^{2}\tilde{m}^{4} n^{6}} & (n\mbox{ odd}),\end{array}\right.
\end{eqnarray}
where $\tilde{\mathcal{C}}^{zz}(n)$ is the ``connected'' correlation function in which the product of magnetizations is subtracted, $\mathcal{C}^{zz}(n) = \left\langle\hat{S}^{z}_{j}\right\rangle  \left\langle\hat{S}^{z}_{j+n}\right\rangle  + \tilde{\mathcal{C}}^{zz}(n)$. Thus, the longitudinal correlation function decays as a power law. Interestingly, this $n^{-6}$ power law has been obtained previously in the continuum limit of this setup~\cite{Iucci2010}. In that work, the corresponding ``mass term'' sine-Gordon model at the solvable Luther-Emery point was suddenly activated, resulting in similar power-law decay of the corresponding two-point correlation functions. 

\begin{figure}
\begin{center}
\includegraphics[width=8.6cm]{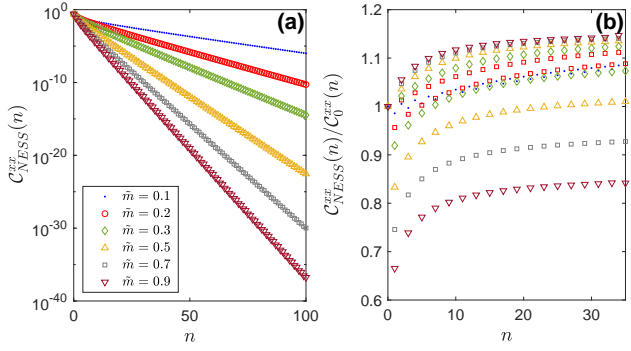}
\caption{(a) Correlation function $\mathcal{C}^{xx}_{\mbox{\scriptsize NESS}}(n) $ computed in the nonequilibrium steady state after a quench from the ground state of the $XY$ model for various values of $\tilde{m} = \frac{m}{J}$. (b) Ratio of correlation $\mathcal{C}^{xx}_{\mbox{\scriptsize NESS}}(n) $ in non-equilibrium steady state to its value in the ground state of Eq.~(\ref{eq:hstag}) shows subleading corrections to purely exponential decay of the ground state.}
\label{fig:cxxhomquench}
\end{center}
\end{figure}

Despite the power-law decay in the longitudinal correlations, the transverse correlations still show exponential decay. Using $B_{j} = c_{j}^{\dagger}-c_{j}$ and $A_{j} = c_{j}^{\dagger} + c_{j}$, one finds
\begin{eqnarray}
\lim_{t\rightarrow \infty}\left\langle A_{j}(t)A_{j+n}(t)\right\rangle & = & \delta_{n=0},\label{eq:homq1}\\
\lim_{t\rightarrow \infty}\left\langle B_{j}(t)A_{j+n}(t)\right\rangle & = & \int_{-\pi}^{\pi}\frac{dk}{2\pi}e^{ikn}\left[\cos\theta_{k}\right.\nonumber\\
& & \left. - (-1)^{j}\sin\theta_{k}\right]|\cos\theta_{k}|,\label{eq:homq2}
\end{eqnarray}
with $\theta_{k}$ defined by Eq.~(\ref{eq:tan1}). One observes that Eqs.~(\ref{eq:homq1})--(\ref{eq:homq2}) are formally similar to the equilibrium result in Eqs.~(\ref{eq:aa0})--(\ref{eq:ba0}) aside from the additional factor of $|\cos\theta_{k}|$ present in Eq.~(\ref{eq:homq2}). As depicted in Fig.~\ref{fig:cxxhomquench}, the decay of the correlation function is exponential with some weak oscillations present in the decay. These oscillations are more easily observed by computing the ratio of the nonequilibrium correlation function to its value in the ground state of the gapped model (c.f., Eq.~(\ref{eq:hstag})) for a particular value of $\tilde{m}$, as shown in the inset. One observes that there are also weak, subleading corrections to the clean exponential decay which occurs in the ground state.

The quench from the ground state of the gapless $XY$ model to the gapped model in Eq.~(\ref{eq:hstag}) most closely resembles the setup in Ref.~\cite{Rieger2018} in which a quench from the gapless phase to the gapped phase of a similar model was investigated. There it was found that the correlation function decayed algebraically $\mathcal{C}^{xx}_{\mbox{\scriptsize NESS}}(n) \sim n^{-\frac{1}{2}}$. This stark contrast in behavior of correlations is likely attributable to several differences between the model presented here and the model investigated in Ref.~\cite{Rieger2018}. First, we employ a constant ``mass term'' proportional to $m$ whereas Rieger et al. employ a staggered field in which the strength satisfies a self-consistency condition. Thus a quench from the gapless ($m=0$) phase to the gapped phase leads to dynamics of the form $m\rightarrow m(t)$ in which the instantaneous mass gap is determined by self-consistency. Such dynamics are entirely absent from the model considered here. Additionally, a constant magnetic field in the $z$ direction is also employed, leading to another adjustable parameter which can be used by Rieger et al. to explore a more complex phase diagram than is needed to describe the model in the present work. While we work exclusively with $h = 0$, the algebraic decay of $\mathcal{C}^{xx}_{\mbox{\scriptsize NESS}}(n) $ observed in Ref.~\cite{Rieger2018} was found with $h\neq 0$ before and after the quench. While Fig.~\ref{fig:cxxhomquench} does show some evidence of weak oscillations in the correlations, Ref.~\cite{Rieger2018} finds that the even and odd correlations decay with different amplitudes in a manner qualitatively similar to what we have found for the domain-wall initial state in which the correlations vanish for odd $n$ and split into separate branches for $\frac{n}{2}$ even or odd. In the case of the domain wall, power-law correlations do persist in the transverse correlation function but with an exponent which is double its value in the ground state of the $XY$ model.
If the final Hamiltonian is instead taken to be Eq.~(\ref{eq:hdim}) corresponding to an energy gap provided by dimerized hopping instead of a staggered magnetic field, similar results are recovered. In addition to the exponential decay in the transverse correlation function, we find power-law decay in the longitudinal correlation function,
\begin{eqnarray}
\mathcal{C}^{zz}_{\mbox{\scriptsize NESS}}(n) &\sim & \left\{\begin{array}{cc} 0 & (n\mbox{ even})\\
\displaystyle -\frac{1}{\pi^{2}\delta^{2}n^{4}} & (n\mbox{ odd}).\end{array}\right.
\end{eqnarray}
In this section, we have computed the long-time limits of observables when the initial, domain-wall magnetization profile is replaced by the ground state of the isotropic $XY$ model. The quench simply consists of suddenly switching on the term corresponding to an energy gap, and we find relaxation of $\mathcal{C}^{xx}_{\mbox{\scriptsize NESS}}(n)$ to exponential decay. However, the longitudinal correlation function $\mathcal{C}^{zz}_{\mbox{\scriptsize NESS}}(n)$ still retains power-law decay. There is some evidence in the literature of power-law decay of longitudinal correlations being more robust than in transverse correlations in domain-wall initial states in which the individual ``halves'' of the system are initially held at different nonzero temperatures~\cite{Aschbacher2003,Aschbacher2006}. The domain-wall magnetization profile investigated in this work is formally a zero-temperature state with the spatial inhomogeneity created by a spatially varying magnetic field, and both transverse and longitudinal correlation functions exhibit power-law decay. 

\begin{figure}
\begin{center}
\includegraphics[width=8.6cm]{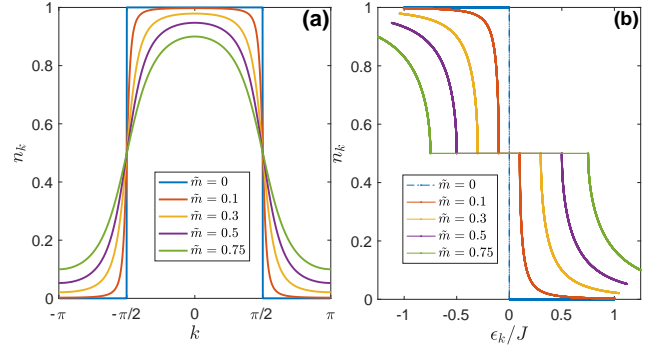}
\caption{(a) Occupation number $n_{k}$ as a function of $k$ for homogeneous initial state. (b) Occupation number $n_{k}$ as a function of mode energy $\epsilon_{k}$ for homogeneous initial state. Viewing $n_{k}$ as an explicit function of $k$ shows how nonzero $\tilde{m}$ mimics a finite temperature by smoothing out the step function in the zero-temperature Fermi sea. The presence of the energy gap is evident when $n_{k}$ is viewed as a function of mode energy, $\epsilon_{k}$.}
\label{fig:specfig}
\end{center}
\end{figure}

The domain-wall magnetization profile is a highly excited initial state, so it is not surprising that power-law decays survive in the presence of an energy gap. The presence of a gap can be expected to only affect the qualitative nature of correlation decay in the ground state or other low-energy settings. However, the ground state of the $XY$ model is also a highly excited state with respect to the Hamiltonian generating time evolution in both cases considered. Taking the staggered field model as an example, we may examine the quasiparticle number $n_{k} \equiv \left\langle \gamma_{k}^{\dagger}\gamma_{k}\right\rangle$ in the initial state using Eq.~(\ref{eq:bog1}),
\begin{eqnarray}
n_{k} & = & \cos^{2}\frac{\theta_{k}}{2}\left\langle c_{k}^{\dagger}c_{k}\right\rangle  + \sin^{2}\frac{\theta_{k}}{2}\left\langle c_{k+\pi}^{\dagger}c_{k+\pi}\right\rangle,\label{eq:nk1}\\
n_{k+\pi} & = & \sin^{2}\frac{\theta_{k}}{2}\left\langle c_{k}^{\dagger}c_{k}\right\rangle  + \cos^{2}\frac{\theta_{k}}{2}\left\langle c_{k+\pi}^{\dagger}c_{k+\pi}\right\rangle\label{eq:nk2}.
\end{eqnarray}
For $|k|<\frac{\pi}{2}$, Eqs.~(\ref{eq:nk1})--(\ref{eq:nk2}) serve to define the quasiparticle occupation numbers for all modes. Using $\left\langle c_{k}^{\dagger}c_{k}\right\rangle = 1$ and $\left\langle c_{k+\pi}^{\dagger}c_{k+\pi}\right\rangle = 0$ for the ground state of the $XY$ model, one finds
\begin{eqnarray}
n_{k} &\rightarrow & \frac{1}{4}\left(1+\frac{\cos k}{\sqrt{\cos^{2} + \tilde{m}^{2}}}\right),\label{eq:nkhom}
\end{eqnarray}
which is valid for all $k \in \left(-\pi,\pi\right)$. The distribution described by Eq.~(\ref{eq:nkhom}) is smooth compared to the zero-temperature Fermi sea in the ground state, $n_{k}^{(0)} = \Theta\left(\frac{\pi}{2}-|k|\right)$. The occupation number is plotted in Fig.~\ref{fig:specfig} for several values of $\tilde{m}$ as a function of $k$ and also as a function of quasiparticle energy, with $\epsilon_{k} = \sqrt{\left(J\cos k\right)^{2} + m^{2}}$. One finds that despite the presence of a gap, high-energy modes in the upper band are already populated in the initial state except at $\tilde{m} = 0$ where the gap vanishes.  

The examination of $n_{k}$ presented here refers explicitly to the homogeneous ground state of the $XY$ model used as the initial state for dynamics generated by $\hat{H}_{s}$ in Eq.~(\ref{eq:hs}). As the operator $\hat{n}_{k} = \gamma_{k}^{\dagger}\gamma_{k}$ commutes with $\hat{H}_{s}$, these mode occupations are conserved and play an important role in the formation of the nonequilibrium steady state. The smoothing of the occupation function in Fig.~\ref{fig:specfig} resembles qualitatively the smoothing that occurs in equilibrium in the presence of a finite temperature. Repackaging this distribution in terms of an effective temperature is rather speculative, as equating Eq.~(\ref{eq:nkhom}) to a Fermi-Dirac distribution with formal ``temperature'' $T(k)$ would introduce a separate temperature for each mode $k$. Such a distribution is quite different from that of thermal equilibrium in which a single temperature is sufficient to capture the entire distribution.

The results in this section emphasize that in addition to the properties of the Hamiltonian which generates the time evolution, the details of the initial state are extremely important to the nature of the non-equilibrium steady state which emerges at long times. The main result of this section is that the persistence of power-law decay in the transverse correlation function appears to be intimately connected to the inhomogeneous, fully-polarized magnetization profile in the initial state. By simply removing the spatial inhomogeneity, the transverse correlation function relaxes to exponential decay. 
\subsection{Comparison to anisotropic $XY$ model}
The most notable result of this work is the persistence of power-law correlations in the long-time limit of the function $\mathcal{C}_{\mbox{\scriptsize NESS}}^{xx}(n)$ despite the existence of an energy gap in the spectrum of the final Hamiltonian. As discussed above, the presence of an energy gap does not necessarily imply that correlation functions away from equilibrium should decay exponentially with distance. However, one does note that the power-law decay of the transverse correlation function changed to exponential decay when the domain wall magnetization profile was replaced by the ground state of the $XY$ model. Here the current-carrying NESS was replaced by a steady state with no spin current. A natural question is how generic these power-law correlations are for current-carrying nonequilibrium steady states in gapped systems.

In this brief section, we note a simple situation in which exponential decay of correlations appears in spite of the existence of a spin current. Consider the anisotropic $XY$ model,
\begin{eqnarray}
\hat{H}_{xy} & = & -J\sum_{j}\left[(1+\gamma)\hat{S}_{j}^{x}\hat{S}_{j+1}^{x} + (1-\gamma)\hat{S}_{j}^{y}\hat{S}_{j+1}^{y}\right] \nonumber\\
& + & h\sum_{j}\hat{S}_{j}^{z},\\
& = & -\frac{J}{2}\sum_{j}\left[c_{j}^{\dagger}c_{j+1} + c_{j+1}^{\dagger}c_{j} + \gamma c_{j}^{\dagger}c_{j+1}^{\dagger} + \gamma c_{j+1}c_{j}\right]\nonumber\\
& + & h\sum_{j}c_{j}^{\dagger}c_{j} + \mbox{constant}.
\end{eqnarray}
In Ref.~\onlinecite{Lancaster2016}, time evolution of observables in this model was considered at a critical point $\gamma = \frac{h}{J} = 1$ where the energy gap vanishes for initial states with a domain-wall magnetization profile. It was shown that at long times, $\mathcal{C}_{\mbox{\scriptsize NESS}}^{xx}(n) \sim \mathcal{M}\cos\left(\frac{\pi n}{2}\right)n^{-\frac{1}{4}}2^{-n}$ as $n\rightarrow\infty$ for some $n$-independent constant $\mathcal{M}$, despite the Hamiltonian generating time evolution being critical (i.e., having no energy gap). The correlation function in the ground state of the anisotropic $XY$ model at this critical point is $\mathcal{C}_{0}^{xx}(n) \sim \mathcal{M}_{0}n^{-\frac{1}{4}}$, so that the non-equilibrium correlations decay much more rapidly with increasing distance than in the ground state. In the present work, we have the {\it opposite} situation in which power-law correlations are persisting in spite of the presence of an energy gap. 

The reason for exponential decay in the $XY$ model at the critical transverse-Ising point is likely due to the extreme mixing of quasiparticles and free fermion operators through a Bogoliubov angle similar to Eqs.~(\ref{eq:bog1}) and (\ref{eq:bog2}) but which mixes creation and destruction operators $\eta_{k} = u_{k}c_{k} + v_{k}c_{v}^{\dagger}$ rather than only mixing creation or destruction operators. The ground state respects this mixing, containing a well-defined number of $\eta$ quasiparticles. However, the domain-wall state will have a well-defined number of $c$ particles but not a fixed number of $\eta$ particles.


To explore how this result is modified as we move away from the critical point, we can use Eq.~(117) from Ref.~\onlinecite{Lancaster2016},
\begin{eqnarray}
\tilde{q}(k) & = & \frac{\mbox{sgn}(k)|\cos k - \frac{h}{J}|}{\sqrt{\left(\cos k - \frac{h}{J}\right)^{2} + \gamma^{2}\sin^{2}k}},
\end{eqnarray}
where $\mathcal{C}^{xx}_{\mbox{\scriptsize NESS}}(n) = \frac{1}{4}\mbox{det}{\bf Q}$ is used to obtain the correlations for general $\gamma$ in which the system is gapped. Here ${\bf Q}$ is a legitimate Toeplitz matrix, so the Fisher-Hartwig conjecture should capture the $n\rightarrow \infty$ asymptotics. The gap in the anisotropic $XY$ model does {\it not} induce a doubling of the unit cell, so this energy gap is of a different qualitative nature than those considered in the present work.

Applying the Fisher-Hartwig conjecture in Eq.~(\ref{eq:fhform}), one finds that at long times
\begin{eqnarray}
\mathcal{C}^{xx}_{\mbox{\scriptsize NESS}}(n)  & \sim & \mathcal{F}(\gamma)\cos\left(\frac{\pi n}{2}\right)(1+\gamma)^{-n},\;\;\;\;(\mbox{as }n\rightarrow\infty)\label{eq:xycor}\nonumber\\
\end{eqnarray}
where $\mathcal{D}(\gamma)$ is a ($\gamma$-dependent) constant. Thus, unlike the cases considered in the present work, the correlations decay exponentially but with the same oscillatory prefactor that is typically associated with the initial state $\left|\Psi_{0}\right\rangle$. Equation~(\ref{eq:xycor}) is a particularly simple example of correlations which decay exponentially in a current-carrying steady state, suggesting that there is no rigorous link between the presence of a current and power-law correlations in gapped models.

\section{Discussion}
\label{sec:conc}
In this work, we computed the long-time behavior of observables in a gapped spin chain after a sudden quench which turns on the energy gap. Our main result is the existence of numerous examples of power-law correlations which exist in a gapped spin chain far from equilibrium. The ground-state correlations of the gapped spin chain generically decay exponentially with distance. However, as we have demonstrated, such power-law correlations can persist in the infinite-time limit when the system begins in a spatially inhomogeneous state which is not an eigenstate of the final Hamiltonian. The two mechanisms for generating an energy gap which have been considered are the application of a staggered magnetic field of constant amplitude $m$ and the dimerization of the $XY$ coupling strength $J$. In both cases, the spin chain Hamiltonian maps to a system of free fermions so that the dynamics may be investigated exactly.

At long times after the quench, both spin-spin correlations decay algebraically rather than exponentially. Interestingly, the domain-wall configuration appears to be quite instrumental in generating the quasi-long-range order, as the transverse correlation function decays exponentially when the ground state of the $XY$ model is used as the initial state instead of the domain-wall state. The longitudinal correlation function, however, retains power-law decay at long times for both types of initial states. It is also interesting to note that the transverse correlation function exhibits exponential decay when the quench is performed with the anisotropic (gapped) $XY$ model as the final Hamiltonian and the initial state possesses a domain wall magnetization profile. The energy gap in the anisotropic $XY$ model arises from a global difference between couplings in the $x$ and $y$ directions. However, both the staggered magnetic field and dimerized hopping break the translation invariance of the system with periodic perturbations to the homogeneous system at the scale of the lattice, doubling the size of the effective unit cell. It is hypothesized that this doubling of the unit cell is closely related to the power-law decay in the transverse correlation function.

All systems considered map to free fermions. A natural question is how robust the results of this paper are with respect to interactions or even weak, integrability-breaking perturbations which are often present in experimental settings. Finely-tuned experiments with cold atoms have been able to verify similar power-law decay and oscillations within correlation functions~\cite{Bloch2015} computed in noninteracting models. Accounting for non-integrable interactions by exact diagonalization is difficult in practice, where only modest system sizes can be handled~\cite{Santos2011}. Compounding this limitation, the energy gap leads to slower motion of the domain wall as shown in Fig.~\ref{fig:j1} so that finite-size effects can influence the results long before the non-equilibrium steady state is reached. 

Many of the exact results regarding the transverse correlation function $\mathcal{C}^{xx}(x)$ both in and away from equilibrium rely on a careful application of the Fisher-Hartwig conjecture, which allows one to extract the asymptotic behavior of Toeplitz matrices as $n\rightarrow \infty$ provided certain conditions are met by the generating function. In this work, we have encountered systems which lead to the evaluation of determinants of matrices which are not of the Toeplitz form, so the Fisher-Hartwig conjecture does not apply. Interestingly, we have been able to obtain partial information about the transverse correlations by identifying a generalized generating function. However, the results are not consistently accurate. For example, the predicted exponential decay factor does not appear to vanish away from equilibrium despite the transverse correlation function decaying as a power law. It is hoped that the results in this work can serve as a test bed for possible generalizations of the Fisher-Hartwig conjecture to matrices with structure which are not exactly of the Toeplitz form.

One potentially promising direction for investigating quench dynamics in the presence of interactions is the application of field-theoretic methods in the continuum limit. Techniques such as bosonization, while rigorously proven to capture the low-energy physics of lattice models, are somewhat uncontrolled approximations away from equilibrium where operators irrelevant to low-energy behavior might strongly influence the dynamics~\cite{Foster2011}. The calculation of nontrivial correlations in exactly solvable models, as presented in this work, provides a benchmark which can be used to calibrate approximate techniques for handling interactions in the continuum limit.

\appendix
\section{Numerical implementation of quench protocol for quadratic Hamiltonians}\label{sec:app1}
In this section we briefly outline a numerical method for investigating quench dynamics when both the initial and final Hamiltonians are quadratic (i.e., noninteracting) in terms of fermionic quasiparticles. Let us consider the time-dependent Hamiltonian
\begin{eqnarray}
\hat{H}(t) = \left\{\begin{array}{cc} \hat{H}_{0} & (t<0)\\ \hat{H}_{f} & (t\geq 0)\end{array}\right.
\end{eqnarray}
At some time $t \ll 0$ the system settles into the ground state $\left|\Psi_{0}\right\rangle$ of $\hat{H}_{0}$. As quadratic models, we may diagonalize the initial and final Hamiltonians in terms of fermionic quasiparticles
\begin{eqnarray}
\hat{H}_{0} & = & \sum_{n}\lambda_{n}\gamma_{n}^{\dagger}\gamma_{n},\nonumber\\
\hat{H}_{f} & = & \sum_{n}\epsilon_{n}\eta_{n}^{\dagger}\eta_{n},
\end{eqnarray}
so that the initial state may be written as
\begin{eqnarray}
\left|\Psi_{0}\right\rangle & = & \prod_{\lambda_{n}\leq 0}\gamma_{n}^{\dagger}\left|0\right\rangle.
\end{eqnarray}
For an observable such as $\hat{O}_{ij} = c_{i}^{\dagger}c_{j}$, one generally wishes to compute $O_{ij} (t) = \left\langle \Psi(t)\right|\hat{O}\left|\Psi(t)\right\rangle$. It is helpful to employ the Heisenberg picture of time evolution so that
\begin{eqnarray}
O_{ij} (t) & = & \left\langle \Psi_{0}\right| \hat{O}(t)\left|\Psi_{0}\right\rangle,\label{eq:numop}
\end{eqnarray}
where $\hat{O}_{ij} (t) = e^{i\hat{H}t}\hat{O}_{ij} e^{-i\hat{H}t}$ with $\hat{H} = \hat{H}_{f}$ for $t\geq 0$. The strategy is to write $\hat{O}_{ij} $ in terms of some combination of the $\gamma_{n}^{\dagger}\gamma_{m}$ operators with the coefficients absorbing the time dependence. In terms of the $\gamma$ operators, the expectation value with respect to the initial state can be computed easily as in the previous section. However, we first have to deal with the time evolution operators $e^{\pm i\hat{H}t}$ by changing to the $\eta$ basis, since these diagonalize $\hat{H}_{f}$. We may assume linear transformations of the form 
\begin{eqnarray}
c_{j} & = & \sum_{m}Z_{jm}\eta_{m} = \sum_{m}V_{jm}\gamma_{m}\nonumber\\
c_{j}^{\dagger} & = & \sum_{m}Z_{jm}^{*}\eta_{m}^{\dagger} =  \sum_{m}V_{jm}^{*}\gamma_{m}^{\dagger},\label{eq:invtr2}
\end{eqnarray}
which provides a basis in which $\hat{H}_{f}$ is represented by a diagonal matrix $D = \mbox{diag}(\epsilon_{n})$
\begin{eqnarray}
\hat{H}_{f} & = & \left(\begin{array}{ccc} c_{1}^{\dagger} & \cdots & c_{N}^{\dagger}\end{array}\right) ZDZ\left(\begin{array}{c} c_{1} \\ \vdots \\ c_{N}\end{array}\right)\\
& = &  \sum_{n}\epsilon_{n}\eta_{n}^{\dagger}\eta_{n}.
\end{eqnarray}
Using this transformation
\begin{eqnarray}
\hat{O}_{ij} (t) & = &  e^{i\hat{H}_{f}t}c_{i}^{\dagger}c_{j}e^{-i\hat{H}_{f}t}\nonumber\\
& = &  e^{i\hat{H}_{f}t}\left[\sum_{m}Z_{im}^{*}\eta_{m}^{\dagger}\right]\left[Z_{jn}\eta_{n}\right]e^{-i\hat{H}_{f}t}.
\end{eqnarray}
Upon acting on $\eta^{\dagger}_{m}$ or $\eta_{n}$, the operator $\hat{H}_{f}$ returns only the corresponding eigenvalue, so
\begin{eqnarray}
\hat{O}_{ij}(t) & = & \sum_{m,n}e^{i\epsilon_{m}t}\left[Z_{im}^{*}\eta_{m}^{\dagger}\right]\left[Z_{jn}\eta_{n}\right]e^{-i\epsilon_{n}t}.
\end{eqnarray}
One can transform to the $\gamma$ basis to calculate the expectation value. This is most straightforwardly accomplished by first transforming back to the $c$ operators and then using Eqs.~(\ref{eq:invtr2}) to transform to the $\eta$ basis:
\begin{eqnarray}
\hat{O}_{ij} (t) & = & \sum_{m,n}e^{i\epsilon_{m}t}\left[Z_{im}^{*}\eta_{m}^{\dagger}\right]\left[Z_{jn}\eta_{n}\right]e^{-i\epsilon_{n}t}\nonumber\\& = & \sum_{k,l}U^{*}_{ik}(t)U_{jl}(t)\gamma_{k}^{\dagger}\gamma_{l}.
\end{eqnarray}
Here the matrix $U_{jl}(t)$ is defined by the ``interior'' sums over the transformations matrices and phase factors. The summations lay out the explicit form for the following matrix multiplication:
\begin{eqnarray}
U(t) & = & Z\Lambda Z^{\dagger}V,
\end{eqnarray}
where
\begin{eqnarray}
\Lambda & = & \left(\begin{array}{ccccc} e^{-i\epsilon_{1}t} & 0 & 0 & \cdots & 0\\ 0 & e^{-i\epsilon_{2}t} & 0 & \cdots & 0\\
0 & 0 & e^{-i\epsilon_{3}t} & \cdots & 0 \\ \vdots & \vdots & \vdots & \ddots & \vdots \\ 0 & 0 & 0 & \cdots & e^{-i\epsilon_{N}t}\end{array}\right).
\end{eqnarray}
Since $U(t)$ is unitary, we have
\begin{eqnarray}
U^{\dagger}(t) & = & V^{\dagger}Z\Lambda^{\dagger}Z^{\dagger},
\end{eqnarray}
and $[U^{\dagger}]_{mn} = U^{*}_{nm}$. Having computed $U(t)$, the time-dependent expectation value is 
\begin{eqnarray}
O_{ij}(t) & = & \left\langle\Psi_{0}\right| \sum_{k}U_{ik}^{*}(t)\gamma_{k}^{\dagger} \sum_{l}U_{jl}(t)\gamma_{l}\left|\Psi_{0}\right\rangle \nonumber\\
& = & \sum_{k,l}U_{ik}^{*}(t)U_{jl}(t)\left\langle\Psi_{0}\right| \gamma_{k}^{\dagger}\gamma_{l}\left|\Psi_{0}\right\rangle\nonumber\\
& = & \sum_{\lambda_{k}\leq 0} U^{*}_{ik}(t)U_{jk}(t),
\end{eqnarray}
where the sum extends only over values of $k$ for which the initial Hamiltonian's eigenvalues are not positive ($\lambda_{k}\leq 0$). Local observables such as magnetization and spin current can be written as a sum of one or two terms in the form of Eq.~(\ref{eq:numop}) and may be computed directly. The function $\mathcal{C}^{zz}(n,t)$, given by Eq.~(\ref{eq:szszness1}), is also a compact expression in terms of the basic contractions. The general form for $\mathcal{C}^{xx}(n,t)$ is a Pfaffian~\cite{Barouch2}, which may be computed directly from a matrix populated by entries of the form in Eq.~(\ref{eq:numop}) using standard libraries~\cite{Wimmer2012}.

\section{General domain walls}\label{sec:app2}
Suppose we consider an initial state $\left|\Psi_{m_{0}}\right\rangle$ constructed by joining two semi-infinite subsystems of uniform magnetization $\pm m_{0}$ where $0\leq m_{0}\leq \frac{1}{2}$, where $m_{0}\rightarrow 0$ corresponds to a homogeneous system with zero magnetization and $m_{0}\rightarrow \frac{1}{2}$ corresponds to the domain walls considered in the rest of this paper
\begin{eqnarray}
\lim_{m_{0}\rightarrow \frac{1}{2}}\left|\Psi_{m_{0}}\right\rangle & = & \left|\Psi_{0}\right\rangle\\
& = & \left|\cdots \uparrow\uparrow\uparrow \downarrow\downarrow\downarrow\cdots\right\rangle.
\end{eqnarray}
Explicitly, $\left|\Psi_{m_{0}}\right\rangle = \left|\Psi_{0}^{L}\right\rangle \otimes \left|\Psi_{0}^{R}\right\rangle$, where
\begin{eqnarray}
\left|\Psi_{0}^{L}\right\rangle & = & \prod_{k=-k_{F}^{+}}^{k_{F}^{+}}L_{k}^{\dagger}\left|0\right\rangle,\;\;\;\;\;\;\;\left|\Psi_{0}^{R}\right\rangle = \prod_{k=-k_{F}^{-}}^{k_{F}^{-}}R_{k}^{\dagger}\left|0\right\rangle,\nonumber\\
\end{eqnarray}
where the $L_{k}^{\dagger}$ ($R_{k}^{\dagger}$) are momentum-basis creation operators like $c_{k}^{\dagger}$ which act only on the left (right) side of the system. Here $k_{F}^{\pm} = \frac{\pi}{2}\pm \pi m_{0}$ represent the effective Fermi momenta on the two homogeneous halves of the system. We can write~\cite{Sabetta2013}
\begin{eqnarray}
\left\langle\Psi_{m_{0}}\right| c_{p+\frac{q}{2}}^{\dagger}c_{p-\frac{q}{2}}\left|\Psi_{m_{0}}\right\rangle & \simeq & \left[\frac{i\Theta(k_{F}^{-}-|p|)}{q+i0^{+}}\right.\nonumber\\
& + &\left. \frac{-i\Theta(k_{F}^{+}-|p|)}{q-i0^{+}}\right].\label{eq:ckckmo}
\end{eqnarray}
Supposing such a state evolves under time evolution generated by the isotropic $XY$ model with a staggered magnetic field [Eq.~(\ref{eq:hstag})], the time evolution of position-space operators is given by Eq.~(\ref{eq:cjtdim}). One may write the time-dependent expectation value of the basic contraction as
\begin{widetext}
\begin{eqnarray}
\left\langle c_{j}^{\dagger}(t)c_{j+n}(t)\right\rangle & = & \int_{-\frac{\pi}{2}}^{\frac{\pi}{2}} \frac{ dp}{2\pi}\int_{-\frac{\pi}{2}}^{\frac{\pi}{2}} \frac{dq}{2\pi}e^{-ipn + iq\left(j+\frac{n}{2}\right)}\left[\right.\nonumber\\
& \times & \left(f^{*}_{p+\frac{q}{2}\, t}f_{p-\frac{q}{2}\,t} + (-1)^{m}g_{p+\frac{q}{2}\, t}^{*}f_{p-\frac{q}{2}\, t} + (-1)^{n}f_{p+\frac{q}{2}\, t}^{*}g_{p-\frac{q}{2}\, t} + (-1)^{m+n}g^{*}_{p+\frac{q}{2}\, t}g_{p-\frac{q}{2}\, t}  \right)\left\langle c_{p+\frac{q}{2}}^{\dagger}c_{p-\frac{q}{2}}\right\rangle \nonumber\\
& + & \left. \left((-1)^{m+n}f_{p+\frac{q}{2}\, t}f^{*}_{p-\frac{q}{2}\,t} + (-1)^{n}g^{*}_{p+\frac{q}{2}\, t}f^{*}_{p-\frac{q}{2}\, t} + (-1)^{m}f_{p+\frac{q}{2}\, t}g_{p-\frac{q}{2}\, t} + g^{*}_{p+\frac{q}{2}\, t}g_{p-\frac{q}{2}\, t}  \right)\left\langle c_{p+\pi+\frac{q}{2}}^{\dagger}c_{p+\pi-\frac{q}{2}}\right\rangle \right]\nonumber\\ \label{eq:bigeq}
\end{eqnarray}
\end{widetext}
We sketch the evaluation of a single term of Eq.~(\ref{eq:bigeq}) in the long-time limit, where the integral over $q$ is dominated by contributions with $q\sim 0$. In this limit the phase $e^{iq\left(j+\frac{n}{2}\right)}$ may be replaced by unity, and
\begin{eqnarray}
f_{p+\frac{q}{2}\, t}^{*}f_{p-\frac{q}{2}\, t} & \simeq & \frac{1}{2}\left(1+\cos^{2}\theta_{p}\right)\cos \left(v_{p} qt\right) + i\cos\theta_{p}\sin\left(v_{p} qt\right),\nonumber\\
\end{eqnarray}
where terms have been dropped which oscillate rapidly as $t\rightarrow \infty$. Here $v_{p} \equiv \partial_{p}\lambda_{p}$ and $\lambda_{p} = \sqrt{(J\cos p)^{2} + m^{2}}$. Changing variables to $u = v_{p}qt$ and taking $t\rightarrow \infty$, the explicit representation of the initial state correlations in Eq.~(\ref{eq:ckckmo}) may be used to write
\begin{widetext}
\begin{eqnarray}
\int_{-\frac{\pi}{2}}^{\frac{\pi}{2}}\frac{dq}{2\pi}f_{p+\frac{q}{2}\, t}^{*}f_{p-\frac{q}{2}\, t}\left\langle c_{p+\frac{q}{2}}^{\dagger}c_{p-\frac{q}{2}}\right\rangle & = & \int_{-\infty}^{\infty} \frac{du}{2\pi}\left[\frac{1}{2}\left(1+\cos^{2}\theta_{p}\right)\cos u + i\cos\theta_{p}\sin u\right)\left[\frac{i\Theta(k_{F}^{-} - |p|)}{u+i0^{+}\mbox{sgn}(v_{p})} + \frac{-i\Theta(k_{F}^{+}-|p|)}{u-i0^{+}\mbox{sgn}(v_{p})}\right].\nonumber\\ \label{eq:contour} \\
& = & \frac{1}{4}\left(1+\cos^{2}\theta_{p}\right)\left[\Theta(k_{F}^{+}-|p|) + \Theta(k_{F}^{-}-|p|)\right] + \frac{1}{2}\mbox{sgn}(p)\left[\Theta(k_{F}^{+}-|p|) -\Theta(k_{F}^{-}-|p|)\right].\nonumber\\
\end{eqnarray}
\end{widetext}
To obtain the last line, we employ the complex exponential representation of the trigonometric functions [e.g., $\cos u = \frac{1}{2}(e^{iu} + e^{-iu})$] and evaluate each term as a contour integral which must be closed on either the upper or lower half of the plane, resulting in one of the two poles being enclosed. The remaining terms in Eq.~(\ref{eq:bigeq}) are evaluated similarly, giving
\begin{eqnarray}
\lim_{t\rightarrow\infty}\left\langle c_{j}^{\dagger}(t)c_{j+n}(t)\right\rangle & = & \left\langle c_{j}^{\dagger}c_{j+n}\right\rangle_{\mbox{\scriptsize NESS}}\\
 & = & \int_{-\pi}^{\pi}\frac{dp}{2\pi}e^{-ipn}G^{(j)}_{m_{0}}(p),
\end{eqnarray}
where
\begin{eqnarray}
 G_{m_{0}}^{(j)}(p) & = & \frac{1}{4}\left(\mathcal{S}_{m_{0}}(p)+\mathcal{S}_{m_{0}}(p+\pi)\right)\nonumber\\
& + & \frac{1}{4}\left(\cos\theta_{p}-(-1)^{j}\sin\theta_{p}\right)\cos\theta_{p}\left(\mathcal{S}_{m_{0}}(p)-\mathcal{S}_{m_{0}}(p+\pi)\right)\nonumber\\
&  + & \frac{1}{2}\left(\cos\theta_{p}\mathcal{D}_{m_{0}}(p)\right.\nonumber\\
& & \left.-\frac{1}{2}(-1)^{j}\sin\theta_{p}\left(\mathcal{D}_{m_{0}}(p) + \mathcal{D}_{m_{0}}(p+\pi)\right)\right)\sigma(p),\label{eq:wigfinal}
\end{eqnarray}
where $\mathcal{S}_{m_{0}}(p) = \left[\Theta(k_{F}^{+}-|p|_{(-\pi,\pi)})+\Theta(k_{F}^{-}-|p|_{(-\pi,\pi)})\right]$, $\mathcal{D}_{m_{0}}(p) = \Theta(k_{F}^{+}-|p|_{(-\pi,\pi)}) - \Theta(k_{F}^{-}-|p|_{(-\pi,\pi)})$, and $\sigma(p)$ is defined in Eq.~(\ref{eq:nessck2}). The function $|\cdot|_{(-\pi,\pi)}$ evaluates the absolute value after mapping the argument to the interval $(-\pi, \pi)$. For example, $\left|-\frac{\pi}{4}\right|_{(-\pi,\pi)}= \frac{\pi}{4}$, while $\left|\frac{\pi}{4}+\pi\right|_{(-\pi,\pi)} = \left|-\frac{3\pi}{4}\right| = \frac{3\pi}{4}$. 

An identical procedure applied to the dimerized $XY$ model yields
\begin{eqnarray}
 G_{\Delta \; m_{0}}^{(j)}(p) & = & \frac{1}{4}\left(\mathcal{S}_{m_{0}}(p)+\mathcal{S}_{m_{0}}(p+\pi)\right)\nonumber\\
& + & \frac{1}{4}\left(\cos\phi_{p}-i(-1)^{j}\sin\phi_{p}\right)\cos\phi_{p}\nonumber\\
& & \times\left(\mathcal{S}_{m_{0}}(p)-\mathcal{S}_{m_{0}}(p+\pi)\right)\nonumber\\
&  + & \frac{1}{2}\left(\cos\phi_{p}\mathcal{D}_{m_{0}}(p)\right.\nonumber\\
& & \left.-\frac{i}{2}(-1)^{j}\sin\phi_{p}\left(\mathcal{D}_{m_{0}}(p) + \mathcal{D}_{m_{0}}(p+\pi)\right)\right)\sigma(p),\nonumber\\ \label{eq:wigfinaldim}
\end{eqnarray}
In the limit $m_{0}\rightarrow \frac{1}{2}$, we have $k_{F}^{+}\rightarrow \pi$ and $k_{F}^{-}\rightarrow 0$ so that $\mathcal{S}_{\frac{1}{2}}(p) = \mathcal{D}_{\frac{1}{2}}(p) = 1$, and Eqs.~(\ref{eq:wigfinal}) and (\ref{eq:wigfinaldim}) reduce to Eqs.~(\ref{eq:nessck1}) and (\ref{eq:gkdim}) in the main text, respectively.

\bibliography{domainwall}
\end{document}